\documentclass[11pt]{article}
\usepackage{amsmath,amsfonts,amssymb,epsfig,color,psfrag}
\usepackage{rotating}
\begin{document}
\title{Explicit Proofs and The Flip}
\author{
Dedicated to Sri Ramakrishna \\ \\
Ketan D. Mulmuley \footnote{Supported by NSF grant CCF-1017760.}
 \\
The University of Chicago}


\maketitle

\newtheorem{prop}{Proposition}[section]
\newtheorem{claim}[prop]{Claim}
\newtheorem{goal}[prop]{Goal}
\newtheorem{theorem}[prop]{Theorem}
\newtheorem{hypo}[prop]{Hypothesis}
\newtheorem{guess}[prop]{Guess}
\newtheorem{problem}[prop]{Problem}
\newtheorem{axiom}[prop]{Axiom}
\newtheorem{question}[prop]{Question}
\newtheorem{remark}[prop]{Remark}
\newtheorem{lemma}[prop]{Lemma}
\newtheorem{claimedlemma}[prop]{Claimed Lemma}
\newtheorem{claimedtheorem}[prop]{Claimed Theorem}
\newtheorem{cor}[prop]{Corollary}
\newtheorem{defn}[prop]{Definition}
\newtheorem{ex}[prop]{Example}
\newtheorem{conj}[prop]{Conjecture}
\newtheorem{obs}[prop]{Observation}
\newtheorem{phyp}[prop]{Positivity Hypothesis}
\newcommand{\bitlength}[1]{\langle #1 \rangle}
\newcommand{\ca}[1]{{\cal #1}}
\newcommand{\floor}[1]{{\lfloor #1 \rfloor}}
\newcommand{\ceil}[1]{{\lceil #1 \rceil}}
\newcommand{\gt}[1]{{\langle  #1 |}}
\newcommand{\C}{\mathbb{C}}
\newcommand{\N}{\mathbb{N}}
\newcommand{\R}{\mathbb{R}}
\newcommand{\Z}{\mathbb{Z}}
\newcommand{\frcgc}[5]{\left(\begin{array}{ll} #1 &  \\ #2 & | #4 \\ #3 & | #5
\end{array}\right)}

\newcommand{\cgc}[6]{\left(\begin{array}{ll} #1 ;& \quad #3\\ #2 ; & \quad #4
\end{array}\right| \left. \begin{array}{l} #5 \\ #6 \end{array} \right)}

\newcommand{\wigner}[6]
{\left(\begin{array}{ll} #1 ;& \quad #3\\ #2 ; & \quad #4
\end{array}\right| \left. \begin{array}{l} #5 \\ #6 \end{array} \right)}

\newcommand{\rcgc}[9]{\left(\begin{array}{ll} #1 & \quad #4\\ #2  & \quad #5
\\ #3 &\quad #6
\end{array}\right| \left. \begin{array}{l} #7 \\ #8 \\#9 \end{array} \right)}

\newcommand{\srcgc}[4]{\left(\begin{array}{ll} #1 & \\ #2 & | #4  \\ #3 & |
\end{array}\right)}

\newcommand{\arr}[2]{\left(\begin{array}{l} #1 \\ #2   \end{array} \right)}
\newcommand{\unshuffle}[1]{\langle #1 \rangle}
\newcommand{\ignore}[1]{}
\newcommand{\f}[2]{{\frac {#1} {#2}}}
\newcommand{\tableau}[5]{
\begin{array}{ccc} 
#1 & #2  &#3 \\
#4 & #5 
\end{array}}
\newcommand{\embed}[1]{{#1}^\phi}
\newcommand{\comb}[1]{{\| {#1} \|}}
\newcommand{\stab}{{\mbox {stab}}}
\newcommand{\perm}{{\mbox {perm}}}
\newcommand{\trace}{{\mbox {trace}}}
\newcommand{\polylog}{{\mbox {polylog}}}
\newcommand{\sign}{{\mbox {sign}}}
\newcommand{\proj}{{\mbox {Proj}}}
\newcommand{\height}{{\mbox {ht}}}
\newcommand{\poly}{{\mbox {poly}}}
\newcommand{\std}{{\mbox {std}}}
\newcommand{\m}{\mbox}
\newcommand{\formula}{{\mbox {Formula}}}
\newcommand{\circuit}{{\mbox {Circuit}}}
\newcommand{\core}{{\mbox {core}}}
\newcommand{\orbit}{{\mbox {orbit}}}
\newcommand{\sie}{{\mbox {sie}}}
\newcommand{\pie}{{\mbox {pie}}}
\newcommand{\cycle}{{\mbox {cycle}}}
\newcommand{\ideal}{{\mbox {ideal}}}
\newcommand{\qed}{{\mbox {Q.E.D.}}}
\newcommand{\proof}{\noindent {\em Proof: }}
\newcommand{\weight}{{\mbox {wt}}}
\newcommand{\tab}{{\mbox {Tab}}}
\newcommand{\level}{{\mbox {level}}}
\newcommand{\vol}{{\mbox {vol}}}
\newcommand{\vect}{{\mbox {Vect}}}
\newcommand{\val}{{\mbox {wt}}}
\newcommand{\sym}{{\mbox {Sym}}}
\newcommand{\convex}{{\mbox {convex}}}
\newcommand{\spec}{{\mbox {spec}}}
\newcommand{\strong}{{\mbox {strong}}}
\newcommand{\adm}{{\mbox {Adm}}}
\newcommand{\eval}{{\mbox {eval}}}
\newcommand{\for}{{\quad \mbox {for}\ }}
\newcommand{\Q}{\mathbb{Q}}
\newcommand{\mand}{{\quad \mbox {and}\ }}
\newcommand{\invlim}{{\mbox {lim}_\leftarrow}}
\newcommand{\directlim}{{\mbox {lim}_\rightarrow}}
\newcommand{\sformal}{{\cal S}^{\mbox f}}
\newcommand{\vformal}{{\cal V}^{\mbox f}}
\newcommand{\crystal}{\mbox{crystal}}
\newcommand{\conje}{\mbox{\bf Conj}}
\newcommand{\graph}{\mbox{graph}}
\newcommand{\ind}{\mbox{index}}

\newcommand{\rank}{\mbox{rank}}
\newcommand{\id}{\mbox{id}}
\newcommand{\str}{\mbox{string}}
\newcommand{\RSK}{\mbox{RSK}}
\newcommand{\wt}{\mbox{wt}}
\setlength{\unitlength}{.75in}

\begin{abstract} 
This article describes a formal  strategy of  geometric complexity theory (GCT) 
to resolve the {\em self referential paradox} in 
the $P$ vs. $NP$ and related problems. The strategy, called the {\em flip},
is to go for {\em   explicit proofs} of these problems. 
By an explicit proof we  mean a proof that
constructs  proof certificates of hardness that are easy to verify, construct and decode.
The main result in this paper says that (1) any proof of the arithmetic implication of the 
$P$ vs. $NP$ conjecture is close to an explicit proof in the sense that it
can be transformed into an  explicit proof
by  proving in addition that arithmetic circuit identity testing 
can be derandomized in a blackbox fashion, and  
(2)  stronger forms of these arithmetic hardness
and derandomization conjectures   together imply  a  polynomial time algorithm for 
a formidable explicit construction problem in algebraic geometry. This may explain
why these conjectures, which look so elementary at the surface, have turned out to be
so hard. 

\end{abstract}

\section{Introduction} \label{sintro}
Geometric complexity theory (GCT) is an approach to fundamental hardness problems in
complexity theory via algebraic geometry and representation theory suggested in 
a series of articles \cite{GCT1}-\cite{GCT8}, which we call 
GCT1-8.  In this article we describe and justify a formal  defining strategy of GCT,
called the {\em flip},
to resolve the {\em self referential paradox} in the $P$ vs. $NP$ and related problems.
This paradox refers to the 
question that is often asked: namely, since   the $P$ vs. $NP$ problem 
is a universal statement about mathematics that says  that
discovery is hard, why  could it  not preclude its own proof and 
hence be independent of the  axioms of set theory?
Resolution of this self referential paradox is generally regarded 
as the root difficulty in this problem; cf.   the survey \cite{aaronson} 
and the references therein.

The flip  strategy of  GCT  to resolve the self referential paradox 
is to go for an {\em explicit proof}.
By an  {\em explicit proof} 
of the nonuniform $P$ vs. $NP$ problem (i.e., $NP \not \subseteq P/poly$ conjecture) 
we essentially mean a proof  that shows existence of  proof certificates
for hardness of an NP-complete function $f(X)=f(x_1,\ldots,x_n)$,
also called {\em obstructions} (to efficient computation of $f(X)$), 
that are {\em short} (of $\poly(n)$ bitsize), {\em easy to verify and construct} 
(in $\poly(n)$ time), and {\em easy to decode}.
By  easy to decode we mean that, given $n$, small 
$m=\poly(n)$, and a short obstruction $s$,
a small set $S_{n,m}(s)= \{X_1,\ldots,X_r\}$, $r=\poly(n)$, of inputs can be constructed in
$\poly(n)$ time such that, for every small circuit $C$ of size $\le m$, 
$S_{n,m}(s)$ contains a counterexample  $X_C$ such that $f(X_C)\not = C(X_C)$.
Here $C(X)$ denotes the function computed by $C$. 
A proof technique that yields  an   explicit proof of the nonuniform
$P$ vs. $NP$ problem   is called a 
{\em flip} (from hard to easy), because in essence it  reduces  the original hardness 
(lower bound) problem to easiness (upper bound) problems: namely, to showing that 
verification, construction and decoding of proof certificates of hardness as per that
technique are easy, i.e., belong to the complexity class $P$. In what sense
this strategy amounts to an   explicit resolution of 
the self referential paradox is explained in Section~\ref{sself}.
See Section~\ref{sflip} for the definition of the flip in the arithmetic setting.

\ignore{This strategy  was formulated after the completion of GCT1 and 2, when it was realized
that these initial papers do not address the self referential paradox.
It was announced briefly without any explanations in \cite{GCThyderabad}. The articles GCT3-5
investigate some basic problems in representation theory motivated by the flip, and the main
result of GCT6, based on GCT1-5 and other results in algebraic geometry and representation theory,
provides an approach to implement the flip in the arithmetic setting 
wherein  the underlying field of computation has characteristic zero; cf. 
Section~\ref{sflip} for the definition of the flip in the arithmetic setting.}


The main results in this article provide a posteriori \footnote{
This strategy  was formulated in a rough form 
after the completion of GCT1 and 2, when it was realized
that these initial papers do not address the self referential paradox.
It was announced briefly without any explanations in \cite{GCThyderabad}. The articles GCT3-5
investigate some basic problems in representation theory motivated by the flip, and the main
result of GCT6, based on GCT1-5 and other results in algebraic geometry and representation theory,
provides an approach to implement the flip in the arithmetic setting 
wherein  the underlying field of computation has characteristic zero.
}
 justification for  this flip strategy.
Specifically, it is  shown (cf. Flip Theorems~\ref{tflip} and \ref{tflip2})  that 
any proof of the arithmetic nonuniform version  of the
$P$ vs. $NP$ conjecture in GCT1 (which is a formal weaker implication of the 
boolean $NP \not \subseteq P/poly$ conjecture) can be converted into 
an  explicit proof 
by  proving in addition that 
circuit identity testing  can be derandomized in a blackbox fashion.
This standard derandomization assumption  \cite{nisan,impa,russell} is 
generally believed to be easier than the target lower bound.
Hence, in this sense, any proof of the arithmetic  $P$ vs. $NP$ conjecture 
is close to an explicit proof.
It is also shown (cf. Flip Theorem~\ref{tflip3}) that 
stronger forms of these arithmetic hardness
and derandomization conjectures together imply a polynomial time algorithm for  a
formidable explicit construction problem in algebraic geometry. 
This may explain why these conjectures in complexity theory, 
which look so elementary at the surface,  
have turned out to be so hard. 

A starting point for the investigation in this article was an analogous 
result (cf. Flip  Lemma~\ref{lflip}) for (weak)
arithmetic hardness of the permanent  that follows easily
from  the hardness vs. randomness principle \cite{impa,russell} and downward
self reducibility of the permanent.  Specifically, it follows by derandomizing the
co-RP algorithm in \cite{impa} for testing if a given arithmetic circuit computes 
the permanent using its downward self reducibility.
But self-reducibility  does not seem to be
as   effective  in the context of the $P$ vs. $NP$ problem, as has already been observed in
other contexts in complexity theory 
(e.g. average vs. worst case hardness \cite{trevisan,fort}). 
The best earlier results in the context of the $P$ vs. $NP$ problem were  proved 
in \cite{atserias,fortnow}. 
Using downward self reducibility, the article  \cite{atserias} gives, 
assuming $NP \not \subseteq P/poly$, a probabilistic 
polynomial time algorithm for finding, given
any small circuit $C$, a counterexample on which it differs from SAT. But this 
algorithm cannot efficiently produce a small set (a proof certificate of hardness)
that contains a counterexample  against 
{\em every} small circuit. The article \cite{fortnow} 
gives under the same assumption a probabilistic polynomial time algorithm with an access to the SAT 
oracle for computing a small set of satisfiable formulae that contains a counterexample against
{\em every}  small circuit claiming to compute SAT. The main difficulty in the 
context of the $P$ vs. $NP$ problem is to accomplish the same task in polynomial time
under reasonable complexity theoretic assumptions without any  access to the SAT oracle.
This difficulty is overcome here in the setting of  the arithmetic  $P$ vs. $NP$ conjecture
using the hardness vs. randomness principle \cite{nisan,impa,russell} 
in conjunction with  characterization by symmetries of a certain  exceptional
function  associated with the complexity class $NP$ in GCT1 (cf. Section~\ref{ssymmetry}). 
Characterization by symmetries is a well known  phenomenon in invariant theory
on which GCT is based. Its crucial role here suggests that it may find 
more applications in complexity theory in future.

The flip lemma (Lemma~\ref{lflip})  for the weak arithmetic hardness of the
permanent also  does not have any direct implications in algebraic geometry, unlike
the stronger flip theorem (Theorem~\ref{tflip3}).
This stronger  theorem  is proved by 
combining the hardness vs. randomness principle and characterization by symmetries with
classical algebraic geometry.

There is also a flip theorem in the boolean setting (Flip Theorem~\ref{tboolean}) for a 
stronger average case form of the boolean 
$NP \not \subseteq P/poly$ conjecture based on the characterization
by symmetries. The main ingradient here is just the formulation of this conjecture.
The rest follows easily from the work \cite{nisan,impa} on derandomization of BPP. 
All the nonuniform  results in this article also have analogues in the uniform setting.

In view of all these results,
the flip strategy of GCT  to go for  
explicit proofs of the $P$ vs. $NP$ and related conjectures
seems  quite  natural.

The rest of this article is organized as follows.
Section~\ref{sarithmetic} describes the arithmetic version of the $P$ vs. $NP$ problem defined
in GCT1.
Section~\ref{sflip} describes the formal flip strategy for resolution of the self referential
paradox via  explicit proofs.
The  flip theorems  in various  arithmetic settings are  stated 
in Section~\ref{sgeneric} and  proved in 
Sections~\ref{ssymmetry}-\ref{sarithproof}. The implication in algebraic geometry 
is pointed out in Section~\ref{simpli}. 
The flip theorem in the boolean setting is stated in Section~\ref{sboolean}. 

No familiarity with algebraic geometry is assumed in this paper. 
The required facts from classical algebraic geometry are only  used as blackboxes. 

\section{Arithmetic versions of the $P$ vs. $NP$ and related problems} \label{sarithmetic}
In this section we recall  the arithmetic version of the $P$ vs. $NP$ problem defined in
GCT1 and also arithmetic versions of the related problems.

\subsection{Arithmetic hardness of the permanent}
By the arithmetic hardness conjecture for the permanent, we mean the problem of showing
that the permanent of an $n\times n$  complex  matrix $X$
cannot be computed by any arithmetic  circuit over $\C$ of $m=\poly(n)$  size,
where by the size of the circuit we mean the total number of nodes in it. 
By the weak arithmetic hardness conjecture, we mean the problem of showing that
the permanent of an $n\times n$ integer matrix $X$
cannot be computed by any arithmetic  circuit  (over $Z$ or $\Q$) of $m=\poly(n)$ total  bit 
size,  where by the total bit size of the circuit we mean the total number of nodes in it plus
the the total bit size of all constants in the circuit. Clearly, the weak arithmetic conjecture
is implied by the arithmetic conjecture. By the strong arithmetic conjecture [cf. GCT1],
we mean the problem of showing that $\perm(X)$, 
the permanent of an $n\times n$ variable 
matrix $X$, cannot be approximated infinitesimally closely by an
arithmetic circuit  over $\C$ of $m=\poly(n)$ size. 
Here by infinitesimally close approximation, we mean
that all coefficients of the polynomial computed by a circuit 
can be made infinitesimally close to that of 
the permanent. Clearly, the strong 
arithmetic conjecture implies the arithmetic conjecture. 

By the arithmetic permanent vs. determinant problem \cite{valiant},
we mean the problem of showing that  $\perm(X)$  cannot be represented 
linearly  as $\det(Y)$, the determinant of 
an $m\times m$ matrix $Y$,
if $m=\poly(n)$, or more generally, $m=2^{\log^a n}$, 
for a fixed constant $a>0$, 
and $n\rightarrow \infty$;  the best known lower bound on $m$ at present is 
quadratic \cite{ressayre}.
Here, by a linear representation, we mean that 
the  entries   of $Y$ are  (possibly nonhomogeneous) 
linear functions (over $\C$) of the entries of $X$. 
The strong arithmetic version of this problem [GCT1] is to show  that 
$\perm(X)$ cannot be approximately infinitesimally closely by an expression of the 
form $\det(Y)$ as above. Clearly, the strong arithmetic version implies that arithmetic 
version. The current best lower bound in the strong arithmetic setting is 
quadratic. It is proved in \cite{landsberg} 
using GCT, and provides the first  concrete lower bound 
application of  GCT  in the context of the permanent vs. determinant problem.
The weak arithmetic version of this problem is to show that 
$\perm(X)$  cannot be represented 
linearly  as $\det(Y)$, where the entries of $Y$ are possibly nonhomogeneous linear functions 
over $\Z$ and the total bit size of the specification of $Y$
is  $\poly(n)$, or more generally, $O(2^{\log^a n})$, 
for a fixed constant $a>0$. Clearly the weak arithmetic version is implied by the arithmetic
version.

A priori, it is not at all  clear that 
the strong arithmetic conjectures above are actually  stronger than
the  arithmetic conjectures.  This is expected  because there are 
functions that can be approximated infinitesimally closely by small circuits 
(of small depth) but conjecturally cannot be computed by small circuits (of small depth);
cf.  Section 4.2 in GCT1.

\subsection{Arithmetic $P$ vs. $NP$ problem} \label{sarithpvsnp}
Next we turn to the arithmetic version of the $P$ vs. $NP$ problem defined in GCT1.
Towards that end, we first associate with the complexity class $NP$ a certain integral function
$E(X)$ that is characterized by its symmetries (cf. Section~\ref{ssymmetry})
like the permanent function
associated with the complexity class $\#P$.

Take a set  $\{ X^j_i | 1 \le j \le k , \: 1\leq i\leq m \}$ of
$m$-dimensional vector variables,
for some fixed constant $k\ge 3$.  Here each $X^j_i$ is an $m$-vector.
So there are $km$ vector variables overall.
Let $X$ be the $m \times k m$ 
variable matrix whose columns consist of these $km$ variable vectors.
For any function $\sigma: \{1,\ldots,m\} \rightarrow 
\{1,\ldots,k\}$, let $\det_\sigma(X)$
 denote the determinant of the matrix $X_\sigma$ whose
$i$-th column is $X^{\sigma(i)}_i$. 
 Define
$E(X)= \prod_\sigma \det_\sigma(X)$
where $\sigma$ ranges over all such functions.  Clearly $E(X)$  is well defined 
over any base field $F$. Let $n=km^2$ be the total number entries in $X$.

The ultimate goal of GCT is:

\begin{conj}[The stronger form of the  $NP \not \subseteq P/poly$ conjecture]
\label{cstrongpvsnp}

Let the base field $F=F_{p}$, $p=q^l$, $q=O(\poly(n))$ a prime  and 
$l=n^a$, for a fixed constant $a>1$.
Then  $E(X)^r$, for any $0< r < p$,   cannot be 
computed by an arithmetic circuit over $F_p$ of $\poly(n)$ size.


\end{conj}

Here the rank $l$ of $F_p$ is required to be large so that 
the size of $F_p$ is much larger than the degree $m k^m$ of $E(X)$.
Computation of $E(X)$ has been conjectured to be hard in this case because to 
decide whether
$E(X)$ is zero over $\Z$ is known to be $NP$-complete (cf. page 451 in \cite{gurvits}).

\begin{prop} (cf. Section~\ref{spstrongpvsnpproof}) \label{pstrongpvsnp}
Conjecture~\ref{cstrongpvsnp}   implies $NP \not \subseteq P/poly$.
\end{prop}

An  intermediate goal is:

\begin{conj} \label{cpvsnpzero} [cf. GCT1]
\ 

\noindent (a) [The (nonuniform) arithmetic 
$P$ vs $NP$ problem] 
Suppose the base field (or ring)  $F=\Q$ or $\C$ (or $\Z$).
Then $E(X)$ cannot be computed by an arithmetic circuit of $\poly(n)$ size over $F$.

\noindent (b) [The weak (nonuniform) arithmetic   $P$ vs. NP problem] 
$E(X)$ (over $\Z$) cannot be computed by an arithmetic circuit over $\Z$ of 
total bit size $O(\poly(n))$.

\noindent (c) [The strong (nonuniform) arithmetic $P$ vs. $NP$ problem] 
$E(X)$  cannot be approximated infinitesimally closely by an arithmetic circuit (over $\C$)
of size $\poly(n)$. 
\end{conj}

Here  (b) over $\Z$  is a weaker implication of  Conjecture~\ref{cstrongpvsnp}
for $E(X)$ over $F_p$. It is also
implied by the usual $NP \not \subseteq P/poly$ conjecture
since, as already remarked, the problem of deciding if $E(X) = 0$ over $\Q$ or $\Z$ 
is NP-complete \cite{gurvits}. Furthermore, (b) is a weaker implication of (a),
because in (a) 
there is no restriction on the bitlengths of the integer constants in the circuit
computing $E(X)$. Only 
the total number of nodes in the circuit needs to be $O(\poly(n))$.
Whereas in (b) the total number of nodes as well as the total bit size of the
constants in the circuit need to be $O(\poly(n))$.

\section{The flip and explicit proofs} \label{sflip}
In this section we describe the formal  flip strategy towards the uniform or 
nonuniform
$P$ vs. $NP$ and related problems in the boolean as well as arithmetic settings.

First let us consider the nonuniform boolean setting. 
Fix an  $NP$-complete function $f(X)=f(x_1,\ldots,x_n)$, say SAT. 
The goal of the nonuniform $P$ vs. $NP$ problem  (i.e., $NP \not \subseteq P/poly$
 conjecture)
is to show that 
there does not exist a small circuit $C$ of size 
$m=\poly(n)$  that computes $f(X)$, $n\rightarrow \infty$. 
Equivalently, the goal is to prove:

\noindent {\bf (HOH: Hard Obstruction Hypothesis)}:
 For every large enough $n$, and $m=\poly(n)$,
there  exists a {\em trivial obstruction
(i.e. a ``proof-certificate'' of hardness)} to the efficient computation of
$f(X)$. Here by a trivial obstruction we mean
a table 
that lists for every small circuit $C$ 
a counterexample  $X$ such that $f(X)\not = C(X)$,
where $C(X)$ denotes the function computed by $C(X)$.

The number of rows of this table is equal
to the number of circuits  of size $m=\poly(n)$.
Thus the size of this table is 
exponential; i.e., $2^{O(\poly(n))}$. 
The time to verify whether a given table is a trivial obstruction
is also exponential,
and so is the time of the obvious algorithm 
to decide if such a table exists for given $n$ and $m$, and to construct one 
if it exists. 
From the complexity theoretic viewpoint, this is a hard
(inefficient) task. So
we call  this trivial, brute force 
strategy for proving the nonuniform $P$ vs. $NP$ conjecture,
based on existence of trivial obstructions,  a {\em hard strategy}--it is 
really just a  restatement of the original problem.
Hence, the terminology  Hard Obstruction Hypothesis.

Any proof strategy for  the  $P$ vs. $NP$ problem has to 
answer the following question: 

\begin{question} 
\label{qfeasible} 
In what sense is the proof strategy  fundamentally 
different from the  trivial, brute force strategy  above and not just an
equivalent reformulation of the original problem?
That is,
in what sense are the proof certificates of hardness (obstructions) of  this
proof strategy  fundamentally better than the trivial obstructions above?
\end{question}

Until this question is answered, however sophisticated a proof strategy may be,
it cannot be considered to be more than a restatement or an equivalent 
reformulation of the original problem.

The most obvious and natural abstract strategy that is
fundamentally better than the trivial strategy is suggested by the
$P$ vs. $NP$ problem itself. Before we define it, 
let us first see what is wrong with the trivial obstruction from the 
complexity-theoretic point of view.
That is quite clear. First, it is long,
i.e.,  its description takes exponential space. Second, it is hard to
verify (and also construct); i.e., it  takes exponential time.
Since $NP$ is the class of problems with ``proof-certificates'' that 
are short (of polynomial-size) and easy to verify (in polynomial-time), 
this then leads to the following strategy
for proving the nonuniform
$P\not = NP$ conjecture, based on proof certificates (obstructions) that are 
short, and  easy to verify (and also easy to construct).
We call this   strategy {\em the flip}: 
from the hard (exponential time verifiable trivial obstructions) 
to the ``easy'' (polynomial time verifiable/constructible new obstructions), and from 
the nonexistence (lower bound problem) to the existence (upper bound
 problem)---existence of
an efficient algorithm to verify and construct an obstruction.

Formally, 
we say that a technique for proving the nonuniform $P \not = NP$ conjecture (using
the function $f(X)$) is
a {\em flip} if 
there exists a family ${\cal O}=\cup_{m,n} {\cal O}_{n,m}$ of bit strings called
{\em obstructions} (or obstruction labels), which serve as 
proof certificates of hardness of $f(X)$,
having  the following  Flip properties   F0-F4.

\noindent {\bf F0 [Short]}:
The set  ${\cal O}_{n,m}$ is nonempty and 
contains a short  obstruction string  $s$ 
if $m$ is small,   i.e., $m=O(\poly(n))$,
or more generally $m=O(2^{\log ^a n})$, $a>1$ a fixed constant. Here
short means the  bitlength $\bitlength{s}$ of $s$ 
is  $\poly(n,m)$. This is $\poly(n)$ if $m=\poly(n)$. 

To state F1, 
we define a {\em small global obstruction set} $S_{n,m}$ to efficient computation of 
$f(X)$, for given $n$ and  $m$, to 
be  a  small set $\{X_1,\ldots,X_l\}$, $l=\poly(n,m)$, of inputs such
that, for any circuit $C$ of size $\le m$, $S_{n,m}$ contains a counterexample
$X_C=X_j$, for some $j\le l$,  such that $f(X_C) \not = C(X_C)$. 
Then:

\noindent {\bf F1 [Easy to decode]: }
Each bit string $s \in  {\cal O}_{n,m}$, $m$ small and $s$ short,
denotes a small global obstruction set $S_{n,m}(s)$ to efficient 
computation of $f(X)$ such that:
(a)  given $s,n$ and $m$, $S_{n,m}(s)$ can be computed in 
$\poly(\bitlength{s},n,m)$ time--in particular, if $s$ is short, 
$S_{n,m}(s)$ can be computed in  $\poly(n,m)$ time--, and (b) given $s,n,m$ and any circuit 
$C$ of size $\le m$, a set $S_{n,m,C}(s) \subseteq S_{n,m}(s)$ of 
$O(1)$ size can be computed in $\poly(\bitlength{s},n,m)$ time such that
$S_{n,m,C}(s)$ contains some counterexample $X_C$ such that 
$f(X_C)\not = C(X_C)$.  
A stronger form of (b) is (b'): given $s,n,m$ and $C$, 
a counterexample  $X_C \in S_{n,m}(s)$ as above can be computed in  
$\poly(\bitlength{s},n,m)$ time (we do not consider it in this paper).


\noindent {\bf F2 [Rich]}: For every $n$ and $m=\poly(n)$, 
${\cal O}_{n,m}$ contains  at least $2^{\Omega(m)}$ 
pairwise disjoint obstructions,
each of  $\poly(n,m)$ bitlength. Here we say that  two obstructions
$s ,s' \in  {\cal O}_{n,m}$ are disjoint if 
$S_{n,m}(s)$ and $S_{n,m}(s')$ are disjoint.

\noindent {\bf F3 [Easy to  verify]:} 
 Given $n,m$ and a string  $s$, 
whether $s$ is a valid   obstruction string  for $n$ and $m$--i.e.,
whether $s \in {\cal O}_{n,m}$--can be verified 
in $\poly(n,\bitlength{s},m)$ time. 
In particular, this time is
$\poly(n)$ when $\bitlength{s}$ and $m$ are $poly(n)$.  

\noindent {\bf F4 [Easy to construct]:} 
For each $n$ and $m=\poly(n)$ a valid  obstruction string
$s_{n,m} \in {\cal O}_{n,m}$ can be constructed in $\poly(n,m)=\poly(n)$ time.

\ignore{
To state F5, we need a definition.
Towards this end, let us assume that the proof technique has 
provided  as in 
the natural proof paper \cite{rudich} 
a formal statement 
of a Useful Property ($UP_n$), which lies at its heart, and assuming which 
it plans to prove $NP \not \subseteq P/poly$. Formally, $UP_n$ is a subset of the set
of all $n$-ary boolean functions. Its definition is 
as in \cite{rudich} and we do not restate it here. 
Let $N=2^n$ be the truth table size of specifying any $n$-ary boolean 
function.

\begin{defn} \label{drigidcomplexity}
We say that a proof technique for the $NP \not \subseteq P/poly$ conjecture 
is 
\begin{enumerate} 
\item 
{\em Nonrigid (or probabilistic)} if  $|UP_n| \ge 2^N/N^c$, for some constant $c>0$,
\item {\em Mildly rigid} if $|UP_n| \le 2^N /N^c$ for every constant $c>0$,
as $n\rightarrow \infty$, or more strongly, 
if  $|UP_n| \le 2^N/2^{n^a}$, for some constant $a\ge 1$,
\item {\em Rigid} if  $ |UP_n| \le 2^{\epsilon N}$,
for some $0 \le \epsilon < 1$, 
\item {\em Strongly rigid} if 
$|UP_n| \le 2^{N^\epsilon}$, for some $0 \le \epsilon < 1$,  and 
\item {\em Extremely rigid} if 
$|UP_n| \le 2^{\poly(n)}$, i.e., $\le 2^{n^a}$ for some constant $a>0$.
\end{enumerate}
\end{defn} 

Thus probabilistic proof techniques to which the natural proof
barrier \cite{rudich} applies are nonrigid, and 
if a proof technique is mildly rigid it bypasses this
barrier, since it violates the largeness criterion in \cite{rudich}. 

With this definition of rigidity, we can now state:

\noindent {\bf F5 [Extremely rigid]:} 
The proof technique is extremely rigid.

The restriction on proofs imposed by this hypothesis  is extremely severe compared to 
the mild rigidity restriction imposed by the natural proof barrier \cite{rudich}.
}
This finishes the description of F0-4 defining a flip. 

We say that a proof of the $NP \not \subseteq P/poly$ conjecture (using $f(X)$) is
{\em extremely  explicit} if it  proves existence 
of an obstruction family ${\cal O}$ satisfying F0-4.
We have defined explicitness in the most extreme 
form here, because we wish to prove the flip results later (Theorems~\ref{tflip2} and 
\ref{tflip3}) in a strongest possible form to indicate what is eventually possible. 
One may also consider weaker  forms of explicitness (as we  do in GCT) by relaxing the
conditions above appropriately. We 
do not define  them here since they are not used in this paper. Hence, in this paper,
whenever we say explicit, we mean extremely explicit.



\subsection{Uniform setting}
Now let us consider the uniform setting.
We say that a technique for proving the uniform $P \not = NP$ conjecture (using
the function $f(X)$) is
a (uniform) {\em flip}, and the resulting proof {\em explicit},  if 
there exists a family ${\cal O}=\cup_{m,n} {\cal O}_{n,m}$ of bit strings called
{\em obstructions} (or obstruction labels), which serve as 
proof certificates of hardness of $f(X)$,
satisfying the Uniform Flip properties  UF0-UF4, which are obtained
from F0-F4 by simply replacing the circuits in their definitions by uniform circuits.
Note that UF1 (b) and UF4 together imply  ``efficient 
diagonalization within $O(1)$ factor'': 
given $n,m=\poly(n)$ and
any  algorithm $C$ that
works within $m$ time on inputs of size $n$, a set 
$S_{n,m,C}$ of 
$O(1)$ size can be computed in $\poly(n,m)$ time such that
$S_{n,m,C}$ contains some counterexample $X_C$ such that 
$f(X_C)\not = C(X_C)$. 


\subsection{Arithmetic setting} \label{sarithflip}
We can similarly define  the flip and explicit proofs for the arithmetic $P$
vs. $NP$ problem (Conjecture~\ref{cpvsnpzero}) letting
$E(X)$ in  Section~\ref{sarithmetic} play the role of $f(X)$.

In the weak arithmetic setting, we 
replace boolean  circuits of bit size $\le m$ by  arithmetic circuits of total bit size $\le m$
in all definitions. 

In the arithmetic setting, we replace boolean  circuits of bit size $\le m$ by 
arithmetic circuits of  size (not bit size) $\le m$
in all definitions. The obstructions in ${\cal O}_{n,m}$ are now meant to be against all
arithmetic circuits of size $\le m$. The running time bounds in 
all the definitions are the same as before except that the running time of the
decoding algorithm in F1 (b) is meant to be $\poly(n,m,\bitlength{s})$, assuming unit-cost
access to the circuit $C$ as an oracle;
the actual cost of evaluating $C$ can be  much larger than $m$ now since there is no 
bound on the sizes of the constants in $C$. 
In the arithmetic setting we will mainly be interested in explicit proofs that 
have the following additional  geometric property $G$. 

To define it, we need some notation.
For given $s \in {\cal O}_{n,m}$, let $S_{n,m}(s)=\{X_1,\ldots,X_l\}$, $l=\poly(n,m)$,
denote the small glbal obstruction set as in F1 (a). Let $V$ denote the space of 
polynomial functions in $X$ of degree $\le 2^m$. Thus the polynomial function $C(X)$ 
computed by any arithmetic circuit $C$ of size $\le m$ belongs to $V$. Let 
$\Sigma=\Sigma_{n,m}=\{C(X)\} \subseteq V$, 
where $C$ ranges over all such circuits. The function  $E(X)$ also belongs to
$V$ assuming that $2^m > \deg(E(X))$. Let $\psi_s: V \rightarrow \C^l$ be the linear map
such that, for any $g(X) \in V$ and any $i \le l$, 
\[ \psi_s(g(X))_i= g(X_i).\] 
In other words, $\psi_s(g(X))$ is simply the $l$-tuple 
of evaluations of $g(X)$ at various $X_i$'s,
and $\psi_s(g(X))_i$ denotes the $i$-th entry in this tuple.
Clearly $\psi_s(E(X)) \not \in \psi_s(\Sigma)$ by the definition of an obstruction.
We call $\psi_s$  an  {\em  explicit linear separator}  associated with $s$.
The geometric property $G$ mentioned above is as follows. 

\noindent {\bf G:} 
The point $\psi_s(E(X))$ does not belong to the closure of 
$\psi_s(\Sigma_{n,m})$ (in the usual 
complex topology) for any $s \in {\cal O}_{n,m}$. 

The motivation here is as follows. In GCT we are interested in showing
existence of an obstruction using algebro-geometric techniques. 
If $\psi_s(E(X))$ belongs to the closure of $\psi_s(\Sigma)$ then any polynomial function 
that vanishes on $\psi_s(\Sigma)$ will also vanish on $\psi(E(X))$. Hence no algebro-geometric
technique will be able to  distinguish $\psi_s(E(X))$ from $\psi_s(\Sigma)$. The 
property $G$  is meant to rule out such pathological  geometric behaviour and ensure 
that the separator $\psi_s$   is good geometrically.

The flip in  the strong arithmetic setting is defined by making the following 
change in the  definitions of F0-4 and G in the arithmetic seting: 
replace a circuit of size $\le m$ (or rather the
function computed by it) everywhere 
by a function that can be approximated infinitesimally closely by circuits of size $\le m$.

We can similarly define the flip and an  explicit proof for the various arithmetic 
versions of the permanent vs. determinant problem, replacing a 
circuit by a linear (determinantal) representation.
We can  also  define these notions 
for other lower bound  problems 
in complexity theory such as the $P$ vs. $NC$ problem.

\subsection{Self-referential paradox} \label{sself}
We now explain in what sense implementation of the flip amounts to 
 explicit resolution
of the self referential paradox, and why this  is  such a  formidable challenge.

Towards this end, let us examine the properties F  above  more closely.
For an  obstruction $s \in {\cal O}_{n,m}$, let  
$S_{n,m}(s)$ denote the corresponding global obstruction set in F1 (a) that 
can be computed in polynomial time. To simplify the argument, 
let us replace F1 (b) by (b)'. The decoding algorithm in (b)' gives in 
polynomial time a counterexample $X_C \in S_{n,m}(s)$ for every small 
circuit $C$ of size $\le m$.
Let $\tilde S_{n,m}(s)$ denote the trivial obstruction of exponential size
that lists for every small
$C$ this $X_C$. Then $S_{n,m}(s)$ can be thought of as a polynomial size 
encoding (i.e., information theoretic compression) of the trivial obstruction 
$\tilde S_{n,m}(s)$.
To verify  a given row of $\tilde S_{n,m}(s)$, we have to check if 
$f(X_C)\not =C(X_C)$ for the $C$ corresponding to that row.
For general $X_C$, this cannot be done in polynomial time, assuming  $P \not = NP$,
since $f$ is $NP$-complete.
And yet F3 says that whether $s$ is a valid obstruction, i.e., whether each of the
{\em exponentially many} rows of $\tilde S_{n,m}(s)$ specifies a counterexample, can be
verified in polynomial time. At the surface, this may seem impossible. It may seem
as if to prove $P\not = NP$, we are trying to prove $P=NP$. 
This is why implementation of the flip is such a formidable challenge.

\ignore{The current state  in complexity theory in the context of the flip
is as follows.
The  proofs of the known  lower bounds (e.g. \cite{hastad}) 
for constant depth circuits and also 
the known quadratic lower bound \cite{ressayre}  for the permanent are 
nonrigid (naturalizable \cite{rudich}) and far from explicit. 
Using characterization by symmetries of the permanent 
(described in Section~\ref{ssymmetry})
and  sparse black box polynomial identity testing 
\cite{agrawal}, it is easy to  give an 
explicit proof for the (trivial)  polynomial lower bound
in the depth two arithmetic circuit model (over $\C$)
for the permanent  (we omit the details). 
But above depth two, the implementation of the flip is a challenge. 
In fact, an  explicit proof of even a very modest linear 
lower bound in the depth three  arithmetic circuit model  is 
a challenge; the proof of the known
quadratic lower bound \cite{amir} in this model is nonrigid (naturalizable) 
and far from explicit. GCTlocal gives a {\em locally explicit}  proof
of a special case of the $P\not = NC$ conjecture, where local explicitness 
is a certain weaker form of explicitness. GCT6 gives 
a  strongly explicit proof of
a nontrivial mathematical form  of the $\#P \not = NC$ conjecture. The proofs 
in GCTlocal and GCT6 are based on nontrivial algebraic geometry, though the
statements of the lower bound results proved therein are elementary.
An approach to implement  the flip in general for characteristic zero 
suggested in GCT6 turns out to be a formidable affair.
}

\section{Main results} \label{sgeneric}
That leads one to ask: why should we then go for 
explicit proofs 
for  the nonuniform $P \not = NP$ and related conjectures 
when just 
proving existence of some obstructions even nonconstructively
suffices in principle?  The reason is provided 
by the following   results (Theorems~\ref{tflip} and \ref{tflip2}) which say that 
any proof of the arithmetic nonuniform 
$P$ vs. $NP$ conjecture (Conjecture~\ref{cpvsnpzero})
can  converted into an  explicit proof 
by proving in addition that arithmetic
circuit identity testing  can be derandomized in a blackbox fashion. This
standard derandomization assumption \cite{impa,russell}
is  generally regarded as easier than the target lower bound.
Hence, in this sense, any proof of the arithmetic  $P$ vs. $NP$ conjecture 
is close to an explicit proof.

\subsection{Weak arithmetic setting}
We begin with a preliminary lemma in the context of the weak 
arithmetic hardness of the permanent as a motivation.

\begin{lemma}[Flip, nonuniform weak arithmetic] \label{lflip}
\ 
Assume the weak arithmetic hardness conjecture for the permanent: specifically, that the
permanent of an $n\times n$ integer matrix $X$
cannot be computed by any arithmetic  circuit (over $\Q$) of $m=\poly(n)$ total 
bit size. 
Suppose also that  the complexity class $E$ (consisting of the problems that
can be solved in exponential time) does not have subexponential
size circuits (or less stringently, that black box 
polynomial identity testing \cite{agrawal,russell}
can be derandomized; cf. Section~\ref{sderandom}). Then:

\noindent (1) For every $n$ and $m=\poly(n)$,
it is possible to compute in $\poly(n,m)=\poly(n)$ time 
a small set $S_{n,m}=\{X_1,\ldots,X_l\}$,  $l=\poly(n,m)=\poly(n)$, 
of $n\times n$ integer matrices such that for every arithmetic circuit 
$C$  of total bit size 
$\le m$, $S_{n,m}$ contains a matrix $X_C$ which is a counter example against
$C$, i.e,  such that $\perm(X_C)$ is not
equal to the value $C(X_C)$ computed by the circuit. 
The set $S_{n,m}$ is thus    a small global obstruction set 
of $\poly(n,m)=\poly(n)$   size against all small circuits of total
bit size $\le m$.

\noindent (2): Furthermore, 
assuming a slight strengthening of the assumption that $E$ does not have subexponential
size circuits (Conjecture~\ref{cstrongassumption1} given later), or less stringently,
that black box polynomial identity testing can be derandomized
(Section~\ref{sderandom}), weak 
arithmetic hardness of the permanent has an  explicit proof. Specifically,
there  exists, for every $n$ and $m=\poly(n)$,  a set 
$\tilde {\cal O}_{n,m}$ of obstructions (bit strings) 
satisfying F0-F4.  

\noindent (3) 
Similar result holds for the weak arithmetic form of the permanent vs. determinant problem 
\cite{valiant} over $\Q$, replacing the second assumption in (1) and (2)
by its weaker version--derandomization of symbolic determinant identity 
testing \cite{russell}.
\end{lemma}

Lemma~\ref{lflip}  follows  (cf. Section~\ref{sproofa}) 
from the hardness vs. randomness principle \cite{impa,russell} 
in conjunction with characterization of the permanent by its symmetries (cf. Section~\ref{ssymmetry}). 
A slightly weaker form of Lemma~\ref{lflip} (everything therein except F1 (b))
follows  easily 
(cf. Section~\ref{sproofa}) by derandomizing  \cite{nisan,impa}
the co-RP algorithm in \cite{russell} for testing 
if a given arithmetic circuit $C$ computes the permanent using its downward self-reducibility.
But we cannot prove an analogous result
in the context of the $P$ vs. $NP$ problem using  self reducibility alone.
Using downward self reducibility, the article  \cite{atserias} gives, 
assuming $NP \not \subseteq P/poly$, a probabilistic 
polynomial time algorithm for finding, given
any small circuit $C$, a counterexample on which it differs from SAT; but this 
algorithm cannot efficiently produce a small {\em global} obstruction set against 
{\em all} small circuits.
The  related article  \cite{fortnow} shows  under the same assumption that
there exists a small global obstruction set of satisfiable formulae which contains,
for every small circuit $C$, a counter example on which it differs from SAT. But 
the algorithm in \cite{fortnow}  for finding this set works in
probabilistic polynomial time assuming access to the SAT oracle. Getting rid of this access
to the SAT oracle is the main problem in the context of the $NP \not \subseteq P/poly$ 
conjecture. 
It is solved in the weak arithmetic setting in the following result.

\begin{theorem}[Flip, nonuniform weak arithmetic] \label{tflip}
Result analogous to the one in Lemma~\ref{lflip}   also  holds for the weak 
arithmetic nonuniform $P$ vs. $NP$ problem 
(cf. Conjecture~\ref{cpvsnpzero} (a)) with 
the integral function $E(X)$ defined 
in Section~\ref{sarithmetic}  playing the role of the permanent in Lemma~\ref{lflip}.
\end{theorem} 

This is proved (cf. Section~\ref{sweakarith}) by combining the hardess vs. randomness 
principle \cite{nisan,impa} with the fact [GCT1] that the function $E(X)$ is 
also {\em characterized  by its symmetries}  just like the permanent (cf. Section~\ref{ssymmetry}).

\subsection{Arithmetic setting} 
We now turn to the  arithmetic setting.

\begin{theorem}[Flip, nonuniform arithmetic] \label{tflip2}

\noindent (a) Assume the strong arithmetic hardness conjecture for the permanent,
and the associated strong derandomization hypothesis 
(defined in Section~\ref{sderanstrong}). 
Then the  strong arithmetic 
hardness conjecture for the permanent has an  
explicit proof having the properties  F0-4 and G. 
If we only assume arithmetic hardness conjecture for the permanent,
and the associated  derandomization hypothesis
(defined in Section~\ref{sderanstrong}), 
then the  arithmetic 
hardness conjecture for the permanent has an 
explicit proof having the properties  F0-4 (but G cannot be guaranteed). 

\noindent (b) Similar results holds for 
the  strong arithmetic $P$ vs. $NP$ and 
 permanent vs. determinant problems (cf. Section~\ref{sarithmetic}).
\end{theorem}

This is proved in Section~\ref{sarithproof} using the 
the hardness vs. 
randomness principle and  the  characterization by symmetries (to prove the properties
F0-4) in conjunction  with some classical algebraic geometry  (to prove the property G).

Unlike Lemma~\ref{lflip} and Theorem~\ref{tflip}, Theorem~\ref{tflip2} has a direct 
implication in algebraic geometry. Specifically, it implies (cf. Theorem~\ref{tflip3}) that 
solutions to the strong arithmetic hardness and derandomization conjectures under
consideration will lead to  polynomial time algorithms for
really  formidable explicit construction problems
in algebraic geometry.

The obstruction family $\tilde {\cal O}$ in  Lemma~\ref{lflip}, or  Theorem~\ref{tflip}
or \ref{tflip2} does not depend on the proof technique at all. 
This  obstruction family  is of no use in 
actually proving hardness of the
permanent or $E(X)$ since the proof 
of its existence assumes this hardness. 
The challenge in the implementation of the flip is to prove existence of  an alternative 
family ${\cal O}$ of obstructions having the  flip properties
{\em without  resorting to any hardness  assumptions}.
The main result of GCT, proved in GCT6, extending the investigation in GCT1-5, gives 
an approach to implement the flip 
for the arithmetic form of the $P$ vs. $NP$ problem (Conjecture~\ref{cpvsnpzero}) and
the permanent vs. determinant problem.

A flip theorem like the one above is meaningful only if the  hardness conjecture 
under consideration is 
harder than the additional derandomization conjecture assumed in its statement. Otherwise,
it will really be talking about the difficulty of this additional
derandomization conjecture.
Thus the flip Theorem~\ref{tflip2}  does  not say anything in the context 
of the quadratic lower bound \cite{ressayre} 
in the permanent vs. determinant problem. Indeed, the known
proof in \cite{ressayre} for this quadratic lower bound is far from explicit.
Here the (analogous) flip theorem  will  talk about the difficulty of the derandomization 
conjecture.


\subsection{Boolean setting} 
Analogue of Theorem~\ref{tflip} also holds in the boolean setting for a  
stronger  average case form of the usual (boolean) $NP \not \subseteq P/poly$
conjecture based on the characterization by symmetries; cf.  Section~\ref{sboolean}.
The main new ingradient here is just formulation of this conjecture. The rest follows easily
from the work \cite{impa} on derandomization of BPP.


\subsection{Uniform setting}
The following results  follow  by uniformizing the proofs
of Lemma~\ref{lflip} and Theorem~\ref{tflip}.

\begin{lemma}[Flip, uniform] \label{lflipu}
Assume  that the permanent of an $n \times n$ integer matrix 
cannot be computed by a uniform circuit of  $m=\poly(n)$ bit size 
and that black box polynomial
identity testing can be derandomized (Section~\ref{sderandom})--this is a uniform assumption.
Then the uniform hardness conjecture under consideration has an  explicit proof 
satisfying   UF0-4; this, in particular, implies 
efficient diagonalization  within $O(1)$ factor. 
\end{lemma}

\begin{theorem}[Flip, uniform] \label{tflipu}
Similar result holds for the weak uniform arithmetic hardness of $E(X)$. 
\end{theorem}

Analogous results also hold in the arithmetic and strong arithmetic settings with 
appropriate definition of uniformity.

\section{Characterization by symmetries} \label{ssymmetry}
We now describe the  phenomenon of characterization by  symmetries on which the proof
of the flip lemma and theorems  are  based.

\subsection{Permanent vs. determinant problem}
In the context of   the permanent vs. determinant problem, this phenomenon is 
that the permanent and  determinant, the functions that are complete and almost 
complete for the complexity classes $\#P$ and $NC$, respectively,  are 
{\em exceptional}, by which we mean
they are   characterized by their symmetries in the 
following sense.

Let $Y$ be an $m \times m$ variable matrix. Then by classical 
representation theory \cite{frobenius} $\det(Y)$ is the unique nonzero
polynomial, up to a constant multiple, 
in the variable entries $y_{ij}$ of $Y$ such that:

\noindent  {\bf (D):}  (1) $\det(A Y^* B)= \det(Y)$,
for any   $A,B \in SL_m(\C)$, 
where $Y^*=Y$ or $Y^t$, and (2) $\det(\lambda Y)=\lambda^m \det(Y)$ for any 
$\lambda \in \C$. 
Thus $\det(Y)$ is  characterized 
by its symmetries, and hence, is  exceptional. We  refer 
to this characteristic property of the determinant as property (D)
henceforth. 

Similarly, let $X$ be an $n \times n$ variable matrix. Then by classical 
representation theory again \cite{marcus}
$\perm(X)$ is the unique nonzero  polynomial, up to a constant multiple,
in the variable entries $x_{ij}$ of $X$  such that 
for any diagonal or permutation matrices $A,B$,

\noindent  {\bf (P):} $\perm(AX^*B)=p(A) \perm(X) p(B)$, 

where $X^*=X$ or $X^t$, and $p(A)$ is defined to be 
the product of diagonal entries, if $A$ is diagonal, and one if $A$ is a permutation
matrix, $p(B)$ being similar. 
Thus $\perm(X)$ is also
characterized   by its symmetries, and hence, is exceptional.
We  refer 
to this characteristic property of the permanent as property (P)
henceforth. In the proof of Lemma~\ref{lflip}, only the property (P) is used.
However, the property (D) is  needed in the GCT approach to the permanent vs. 
determinant problem; see the overview \cite{GCTcomplexity}.

For convenience, we now recall the elementary proof of property (P) \cite{marcus}, 
the proof of property (D) being similar.
Let $f(X)$ be any polynomial with property (P). Letting 
$A$ and $B$ in (P) be diagonal matrices, it easily follows that $f(X)$ has the same
total degree as $\perm(X)$, and also the same total degree (one) in the variables of
any fixed row or column of $X$. This means that each monomial of $f(X)$ contains
precisely one variable (with degree one) from each row and column of $X$. Thus 
it corresponds to a permutation of $n$ symbols.
Furthermore,  letting $A$ and $B$ in (P)  be permutation matrices, it follows 
that the coefficients of all monomials are the same. Hence $f(X)$ is a constant
multiple of $\perm(X)$. This proves property  (P).

\subsection{Arithmetic $P$ vs. $NP$ problem} 
The  function $E(X)$ (cf. Section~\ref{sarithmetic}) which plays the 
role of the permanent in the $P$ vs. $NP$ problem  is also  characterized by its
symmetries (Theorem~\ref{tE}).

To state the result, we follow the same notation as in Section~\ref{sarithpvsnp}. 
Let $K$ be the wreath product of the symmetric group $S_k$ on $k$ letters  and
the alternating group $A_m$ on $m$ letters.
It acts on $X$ by permuting its columns in the obvious way.
We call $X_{\sigma_0}$, where $\sigma_0(i)=1$ for all $i$,
the {\em primary submatrix} of $X$, and $\det_{\sigma_0}(X)=\det(X_{\sigma_0})$
the primary minor of $X$.

The following is a strengthening of Proposition 7.2 in GCT1.

\begin{theorem} \label{tE}
Let the base field $F$ be of characteristic zero, say $\Q$ or $\C$. Then:

\noindent (a) $E(X)$ is the only nonzero polynomial, up to a constant multiple,
in the variable entries of $X$ such that

\noindent {\bf (E)}: 

\begin{enumerate} 
\item[E1:]  for any $A \in GL_n(\C)$ and any $B \in K$, 
$E(A X B)= (\det(A))^{k^m} E(X)$.

\item[E2:] (1) $E(X)=0$  for any $X$ with  singular primary minor, or less stringently,
(2) $E(X)=0$ for any $X$ whose primary minor has a unit $(n-1)\times (n-1)$ matrix as 
its top-left $(n-1)\times (n-1)$ minor and  zeros in the bottom row. 

\end{enumerate}

\ignore{\noindent (b) Any nonzero polynomial $e(X)$ satisfying E1
can be written as  a linear combination $\sum_{\alpha} a(\alpha) g(\alpha)$, $a(\alpha)
\in \C$, where $\alpha$ ranges over  monomials
in the $m\times m$ minors of $X$ of the same degree as $E(X)$, and
\[ g(\alpha)=\sum_{B\in K} \alpha(X B).\] }

\noindent (b) Let $e(X)$ be any integral nonzero polynomial satisfying E2 and the following variant of E1:

E1': for any $A \in SL_n(\C)$ and any $B \in K$, $e(A X B)= e(X)$.

Then $e(X)$ can be written as  $E(X)(\sum_{\alpha} a(\alpha) g(\alpha))$, $a(\alpha)
\in \C$, where $\alpha$ ranges over  monomials
in the $m\times m$ minors of $X$,  and
\[ g(\alpha)=\sum_{B\in K} \alpha(X B).\] 
\end{theorem}

We refer to the characterization of $E(X)$ in characteristic zero
given by this result as property (E) henceforth. 

\proof 

\noindent (a) Let $f(X)$ be any polynomial over $\Q$ or $\C$ with property (E).
It is easy to see that E1 and E2 (2) together imply E2 (1). Hence, let us 
assume that $f(X)$ has properties E1 and E2 (1).

By E2 (1),  $f(X)=0$ if the primary minor of $X$ is singular.
Hence it easily follows from Hilbert's Nullstellansatz \cite{mumford} 
that $f(X)$ is divisible by $\det(X_{\sigma_0})$,
where $X_{\sigma_0}$ denotes the primary $m\times m$ minor of $X$. 
Specifically, 
let ${\cal X}$ be the variety consisting of 
$X$'s with singular primary minors. It is the zero set of the polynomial
$\det(X_{\sigma_0})$. 
By E2 (1),  $f(X)$ vanishes on ${\cal X}$. Hence, it follows from 
Hilbert's Nullstellansatz that $f(X)^r$, for some positive integer $r$, is 
divisible by $\det(X_{\sigma_0})$. Since $\det(X_{\sigma_0})$ is irreducible,
it follows that $\det(X_{\sigma_0})$ divides $f(X)$.

\noindent {\em Remark:} The above special case of Nullstellansatz has an 
elementary proof. Specifically, let $\tilde X \in {\cal X}$ be a ``generic'' matrix 
with singular primary minor. Here generic means all entries of $\tilde X$ are 
algebraically independent except (say) the top-left, which is a rational function
of the remaining entries of $\tilde X$ in such a way that the determinant of the
primary minor of $\tilde X$ is zero. Then since $f(X)$ vanishes on
$\tilde X$ and  the determinant is irreducible, it is easy to show
that $\det(X_{\sigma_0})$ divides $f(X)$.

Since, by E1,  $f(X B)=f(X)$ for every $B \in 
K$, it now follows that $f(X)$ is divisible by $\det_\sigma(X)$, for every $\sigma$.
That is, $f(X)$ is divisible by $E(X)$. 
It follows from  E1, by letting $A=\lambda I \in GL_n(\C)$, that
$f(\lambda X)=\lambda^d f(X)$, for any $\lambda \in \C$, 
where $d=m k^m$ is the degree of $E(X)$. This means $f(X)$ is a homogeneous polynomial 
of the same degree as $E(X)$ and is divisible by $E(X)$.
Hence, it is a constant multiple of $E(X)$.  This proves (a).

\noindent (b) Now suppose that $e(X)$ is any nonzero polynomial satisfying E1' and E2.
It follows as above that $e(X)$ is divisible by $E(X)$.
By E1', $e(A X) = e(X)$ for any $A \in SL_n(\C)$. Hence, 
by the  first fundamental theorem of invariant theory \cite{fultonrepr,weyl},
$e(X)$ can be written as a polynomial in the  $m\times m$  minors of $X$. 
Since $e (X B)=e(X)$ for any $B \in K$ and $E(X)$ divides $e(X)$, (b)
follows. \qed

\ignore{
The obstruction family $\tilde {\cal O}$ in the flip Theorem~\ref{tgeneric}
does not depend on the proof technique at all. 
This  obstruction family  is of no use in 
actually proving hardness of the
permanent or $E(X)$ since the proof 
of its existence assumes this hardness.
In other words, this  proof is circular.

The actual  proof  has to come up with an alternative 
family ${\cal O}$ of obstructions satisfying the  flip hypotheses 
{\em without  resorting to any hardness  assumptions}. 
This breaking of the circle (of equivalences), 
that is implementing the flip without resorting to any (equivalent)
hardness assumptions,  is  the main complexity theoretic 
challenge in the fundamental lower bound problems of  complexity theory.  
}

\ignore{
\begin{figure} 
\begin{center}
\epsfig{file=trivialobs.eps, scale=.5}
\end{center}
      \caption{Trivial obstruction}
      \label{fig:trivialobs}
\end{figure} 
}

\ignore{It should also be stressed  that the explicit proof 
barrier applies only to  hardness problems 
that are  harder than derandomization of polynomial identity testing
(which is a lower bound problem in disguise \cite{russell}). Because, as already
remarked, the  flip theorem is meaningful only if  the first
hardness assumption in its statement is  harder than 
the second hardness assumption (derandomization of polynomial identity testing).
Otherwise, it would be talking about the difficulty of the second hardness assumption
rather than the first.
Thus the proof of the  flip theorem 
does not say anything in the context of the quadratic 
lower bound for the permanent, which is much easier than the derandomization problem
and has a nonrigid (naturalizable) nonexplicit proof \cite{ressayre}, or similarly,
in the context of the quadratic lower bound \cite{amir} for
depth three arithmetic circuits, and so forth.}

\section{The stronger form of the $NP \not \subseteq P/poly$ conjecture} 
\label{spstrongpvsnpproof}
Before turning to the proof of the flip lemma and theorems, we prove in this section
Proposition~\ref{pstrongpvsnp} following the same notation as
in Conjecture~\ref{cstrongpvsnp}.

Let $\sigma: F_p \rightarrow F_p$ be the Frobenius automorphism $x \rightarrow x^q$. 
For $x \in F_p$, let  
\[\trace(x)= \sum_{i=0}^{l-1} \sigma^i(x)= \sum_{i=0}^{l-1} x^{q^i}\]
denote its trace.
It is known (Theorem 5.2 in chapter 6 in \cite{lang})
that the bilinear form $\trace(x y)$, $x,y \in F_p$,
is nondegenerate. Fix a basis $B=\{b_i\}$, $0 \le i \le l-1$, of $F_p$ over $F_q$.
Let $\{b_i^*\}$ denote its dual basis with respect to the trace form.
For any $x \in F_p$, let $x_i$'s denote its coefficients in the basis $B$.
Then $x_i=\trace(b_i^* x)$. Hence, for any fixed $i$, 
$x_i \in F_q$ can be computed by an arithmetic $F_p$-circuit (with input $x$)
of $O(l^2)=\poly(n)$ size.
Furthermore, since $q=\poly(n)$, a bit representation 
of $x_i$ can be computed by an $F_q$-circuit of $\poly(n)$ size using Lagrange 
interpolation. Thus, given $x\in F_p$, all bits of all $x_i$'s can be computed 
by an arithmetic $F_p$-circuit of $\poly(n)$ size. 

Now let $e(X)=E(X)^{p-1}$. 
Then $e(X)$ is $1$ iff 
$E(X)$ is nonzero, and it is zero otherwise. Thus $e(X)$ is a boolean function that 
belongs to co-NP. 
So to prove the usual nonuniform $P \not = NP$ 
conjecture over the boolean field, it suffices to show that $e(X)$
can not be computed by a boolean circuit of $\poly(n)$ size, when
the input to the circuit consists of the bits of the 
coefficients of $X_{ij}$ (the entries of $X$) with respect
to the basis $B$.
Suppose to the contrary that such a circuit $C$ exists. Then using $C$ we can construct
an arithmetic circuit $C'$ over $F_p$ of polynomial size computing $e(X)$. Specifically,
we  compute the bits of the 
coefficients of $X_{ij}$ with respect to the basis $B$
by small circuits as above and then feed these bits to $C$. By
Conjecture~\ref{cstrongpvsnp}  such a small $C'$  for computing $e(X)=E(X)^{p-1}$ 
cannot exist. A contradiction.  This proves Proposition~\ref{pstrongpvsnp}. 

\section{Flip in the weak arithmetic setting} \label{sweakarith}
In this section we prove Lemma~\ref{lflip} and Theorem~\ref{tflip}.

\subsection{Proof of Lemma~\ref{lflip}} \label{sproofa}

\begin{prop}\cite{russell}  \label{prussell}
The problem of deciding if a given arithmetic circuit $C$ over $\Z$ 
computes the permanent belongs to
$co-RP$. 
\end{prop} 
\noindent {\em Original proof:} 
We first recall a proof from \cite{russell} and then give a new
proof based on the property (P) that  is crucially needed for  proving F1 (b). 

Given a circuit $C=C_n$ that is supposed to compute $\perm(X)$, $\dim(X)=n$, 
we get a circuit $C_i$, $1 \le i \le n$, for computing $\perm(Y)$, $\dim(Y)=i$,
by putting $Y$ in the lower right corner of $X$, specializing the remaining 
diagonal entries of $X$ to $1$, all others remaining entries to zero, and 
evaluating $C_n$ on this $X$.  

Let $C_i(Y)$ denote the value computed by $C_i$ on input $Y$.
Then $C_n$ computes $\perm(X)$ if and only if for all $1< i \le n$

\begin{equation} \label{eqperm1}
 C_i(Y)=\sum_{j=1}^i y_{1,j} C_{i-1}(Y_j),
\end{equation}
where $Y$ is an $i\times i$ variable matrix with variables $y_{k,l}$, 
and $Y_j$ the $j$-th minor of $Y$ along the  first row,
and 
\begin{equation} \label{eqperm2}
C_1(y)=y.
\end{equation}
This is the usual downward self reducibility of the permanent.
Testing  if $C_i$'s satisfy  (\ref{eqperm1}) and (\ref{eqperm2}) is an 
arithmetic circuit (polynomial) identity testing 
problem (over $\Z$), which belongs to co-RP \cite{ibarra}.

\noindent {\em New GCT proof}:  By the property (P), $C(X)=\perm(X)$, up to a nonzero
constant multiple, if and only if

\begin{equation} \label{eqpropp0} 
C(X) \not =0, 
\end{equation}
identically as a polynomial, 
\begin{equation} \label{eqpropp1}
C(e_i X)=C(X) \mbox{\ and \ }  C(X e_i)=C(X), \quad \mbox{for all } i < n,
\end{equation}
where $e_i$ denotes an elementary permutation matrix (which permutes the $i$th and
$(i+1)$-st positions), and

\begin{equation} \label{eqpropp2}
C(\mu X)= p(\mu) C(X) \mbox{\ and \ }  C(X \mu)=p(\mu) C(X),
\end{equation}
where $\mu$ denotes a  diagonal matrix, and $p(\mu)$ is the product of its 
diagonal entries.

Testing if $C(X)$ satisfies (\ref{eqpropp0})--(\ref{eqpropp2}) is
again an arithmetic circuit identity testing problem over $\Z$,
which belongs to co-RP. \qed

\subsubsection{Proof of Lemma~\ref{lflip}  (1)}
Now consider the second (new)  co-RP algorithm in the proof above to test if
$C$ computes $\perm(X)$. This algorithm works in expected  time $\le m'=m^c$, where
$c>1$ is some fixed constant. Assuming $E$ does not have subexponential size 
circuits, it can be derandomized as follows. Article \cite{impa} gives, under this
assumption, a $\poly(n,m)$ time computable 
pseudorandom generator $g$ that takes  a random seed of $l=O(\log m)$ bit size 
and produces a pseudorandom sequence of length $m^c$
that fools any small circuit of  bit size $\le m^c$.
Consider the computational circuit corresponding to the above co-RP algorithm 
for testing if $C$ computes $\perm(X)$. Feeding the pseudorandom sequence 
generated by $g$ to this circuit in place of the random bits, cycling over 
all $\poly(m)$ possible seeds, and then taking a majority vote,
we get a $\poly(n,m)$  time algorithm $A$ for 
testing if $C$ computes $\perm(X)$.  (For this argument, we only need
derandomization of polynomial identity testing, instead of the strong assumption 
that $E$ does not have subexponential size circuits; cf. Section~\ref{sderandom}
for further discussion.) 

A crucial property of $A$ is that it is {\em nonadaptive}. This means 
the queries generated during its execution do not depend on $C$ at all.
Here a  query specifies an $X$ on which (\ref{eqpropp0}) is tested, or 
an $X$ and $i$ on which  the 
equation (\ref{eqpropp1}) is  tested, or a 
$\mu$ and an $X$ on which (\ref{eqpropp2}) is tested.
Let $Q_{n,m}$ denote the set of $\poly(n,m)$ queries generated in $A$ when the
input to $A$ is a circuit 
$C$ of bit size $\le m$. Nonadaptiveness means $Q_{n,m}$ depends only on $n$ and $m$
but not on $C$ at all.

Assuming that $\perm(X)$ cannot be computed by a circuit of bit size $m=\poly(n)$, 
it follows that, when $m=\poly(n)$,  then for {\em every} $C$ of bit size $\le m$,
$Q_{n,m}$ contains a query 
on which an algebraic identity test based on  (\ref{eqpropp0}),
(\ref{eqpropp1}) or (\ref{eqpropp2})  fails for that $C$. 
Let $S_{n,m}$ be the set of all inputs $X$'s on which $C$ is evaluated 
during the testing of all queries in $Q_{n,m}$. Specifically, fix a query $q$ in 
$Q_{n,m}$. Suppose this query requires testing of the first equation in 
(\ref{eqpropp1}) for some fixed 
$i<n$ and $X=X_q$ for some input $X_q$, the argument for the second equation being
similar.
Then during the course of testing this equation 
for this query, we evaluate 
$C$ on $X_q$ as well $e_i X_q$ 
(The evaluation in the co-RP algorithm \cite{ibarra}
for algebraic identity testing works modulo a large enough prime to keep the bit sizes
under control. But this makes no difference in the argument that follows.)
So there are 
two values of $X$ (namely $X_q$ and $e_i X_q$)  on which $C$ is evaluated 
during the testing of this query. Let $S_q=\{X_q, e_i X_q\}$ and 
add both elements in $S_q$ to $S_{n,m}$ for this
query. If the query $q$ requires testing of the first (say) equation
in (\ref{eqpropp2}) on some fixed value $\mu_q$ of
$\mu$ and $X_q$ of $X$, then we let $S_q=\{X_q, \mu_q X_q\}$, 
add both $X_q$ and $\mu_q X_q$ to $S_{n,m}$. If the query requires testing 
of (\ref{eqpropp0}) on some $X_q$, we let $S_q=\{X_q\}$, and add $X_q$ to $S_{n,m}$.
Thus $S_{n,m}=\cup_q S_q$ contains a set of $\poly(n,m)$ $n \times n$ matrices. 
Because $Q_{n,m}$  contains, for {\em every} $C$ of bit size $\le m$, a query 
on which the associated algebraic identity test fails, it follows that 
$S_{n,m}$ also  contains, for {\em every} circuit $C$ of bit size $\le m$, 
a matrix $X_C$ on which $C(X_C) \not = \perm(X_C)$. Thus $S_{n,m}$ is
a small global obstruction set against all circuits of bit size $\le m$. 
Furthermore, using the algorithm
$A$, we can compute $S_{n,m}$ in $\poly(n,m)$ time. 
This proves   statement (1)  of Lemma~\ref{lflip}.

\subsubsection{Proof of Lemma~\ref{lflip} (2)}
Now we turn to the construction of the obstruction family 
$\tilde {\cal O}=\tilde {\cal O}_{n,m}$ as  needed in the statement (2) of
Lemma~\ref{lflip}. 
Let $m'=m^c$ be  the bound on the running time of $A$ as above. 
Let $l=b \log m$, for a large enough constant $b>c$. For small   $m$ 
(i.e. $m=\poly(n)$), 
let ${\cal O}_{n,m}$ be the set of all $(\log m', a \log m)$-designs within
the set $\{1,\ldots,l\}$, for a 
large enough constant $c<a < b$. Here by a $(k,r)$-design within
$\{1,\ldots,l\}$, we mean \cite{nisan} 
a collection of sets $\{T_1,\ldots,T_{m'}\}$, 
$T_i \subseteq \{1,\ldots,l\}$, such that (1) for all $i$, $|T_i|=r$, and 
(2) for all $i\not = j$, $|T_i \cap T_j|\le k$.
Each such design $s$ 
can be specified by an $m'\times l$ boolean adjacency matrix whose 
$i$-th row  specifies $T_i$ (by letting its $j$-th  entry be one if $T_i$ contains $j$
and zero otherwise). The bitlength $\bitlength{s}$ of this specification is
 $O(m' \log m)=O(\poly(n,m))$. This  $s$ in ${\cal O}_{n,m}$ is short if $m$ is small.
It is easy to see (from the proof of Lemma 2.6 in \cite{nisan}) that
the total number of such designs is $ \ge 2^{\Omega(m' l)}=2^{\Omega(m' \log m)}$. 

We now verify that this construction satisfies F0-F4. For the proof of F2 
we will need a complexity theoretic conjecture.

\noindent {\bf F0:} This is clear by the preceding remark on the number of designs.

\noindent {\bf F1 (a):} It follows from the results in \cite{impa} that, for 
each design $s \in {\cal O}_{n,m}$,  there exists a $\poly(m)$-time computable 
pseudo-random generator $g(s)$ that takes a random seed of $l=O(\log m)$ bit length and 
produces a pseudorandom sequence of bit length $m'=m^c$ that fools any circuit 
of bit size $\le m'$. 
When $m$ is small (and thus $s$ is short),
using this pseudo-random generator $g(s)$ in place of the
pseudo-random generator $g$ above,
we can compute a small global obstruction set $S_{n,m}(s)$ in $\poly(n,m)$ time, so also
the associated set $Q_{n,m}(s)$ of queries. This proves F1 (a).

\noindent {\bf  F1 (b):} Given $n,m=\poly(n)$, a short $s$,
and a circuit $C$, $Q_{n,m}(s)$
is guaranteed
to contain a query $q$ on which $C$ fails, and this query $q$ can be computed in
$\poly(n,m,\bitlength{s})=\poly(n,m)$ time. Let $S_q(s)$ be the associated set
of $X$'s on which $C$ is evaluated during the testing of this query. 
The size of $S_q(s) \le 2$. Let $S_{n,m,C}(s)=S_q(s)$. Clearly it too can be
computed in $\poly(n,m)$ time.

\noindent {\bf F3:} Given a design $s \in {\cal O}_{n,m}$ specified as 
an $m' \times l$ adjacency matrix, whether it is
a valid $(\log m', a \log m)$ design within $\{1, \ldots, l\}$, $l=b \log m$, 
can be clearly verified in $\poly(n,m)$ time.

\noindent {\bf F4:} 
Lemma 2.6 in \cite{nisan} gives an algorithm  to compute one such 
valid design in $\poly(n,m)$ time. 

\noindent {\bf F2:}  
This follows from:

\begin{conj} \label{cstrongassumption1}
The pseudorandom generator 
$g(s)$ given by \cite{impa} under the assumption
that $E$ does not have subexponential size circuits has the following
additional property:
for a fixed constant $c$,
and  large enough constants $a>c$ and $b>a$, 
${\cal O}_{n,m}$ contains  $2^{\Omega(m)}$ mutually disjoint $s$'s 
(as we would expect if  $s$'s are sufficiently (pseudo)-random).
Here we say that 
$s,s'$ are mutually disjoint if $S_{n,m}(s)$ and $S_{n,m}(s')$ are mutually 
disjoint.
\end{conj} 

This is a slightly strengthened version of the following conjecture 
that only depends on the complexity class $E$, and not on the permanent vs. 
determinant problem or the property (P).

Let $R_m(s)$ 
denote the set of pseudorandom sequences of length $m^c$  produced by $g(s)$ as 
the seed ranges over all possible bit-strings of length $l=b \log m$. 

\begin{conj} \label{cstrongassumption1'}
The pseudorandom generator 
$g(s)$ given by \cite{impa} under the assumption
that $E$ does not have subexponential size circuits has the following additional property:
for a fixed constant $c$,
and  large enough constants $a>c$ and  $b>a$, the collection 
$\{R_m(s)\}$, $s \in {\cal O}_{n,m}$,  contains 
at least $\Omega(2^{\Omega(m)})$ mutually disjoint sets.
\end{conj}

Each string in $R_m(s)$ contributes $\poly(n)$ $X$'s to $S_{n,m}(s)$,
instead of just one,
and hence disjointness of $S_{n,m}(s)$'s in Conjecture~\ref{cstrongassumption1} 
is a bit stronger than disjointness of  $R_m(s)$'s  above.
Conjectures~\ref{cstrongassumption1} and \ref{cstrongassumption1'}  stipulate 
pseudo-randomness of the generator in \cite{impa} with respect to a new measure 
in addition to the usual one used there.

\subsubsection{Proof of Lemma~\ref{lflip} (3)}
This is similar to that of Lemma~\ref{lflip} (1) and (2).

This finishes the proof of Lemma~\ref{lflip}. 

\subsection{Characterization by symmetries vs. self reducibility} \label{sselfred}
It is illuminating to consider what happens if we use in the preceding proof
the first (original) co-RP  
algorithm in the proof of Proposition~\ref{prussell} instead of the second (new) one
as we did.
Then  we cannot prove F1 (b).  Because each query to test
(\ref{eqperm1}) in Proposition~\ref{prussell} 
requires $O(n)$ evaluations of the circuit $C$. Hence
the size of $S_{n,m,C}(s)$ in this case would be $O(n)$ and not $O(1)$
as needed in F1 (b).  Thus  the new co-RP algorithm is crucial to bring down
the size of $S_{n,m,C}(s)$ from $O(n)$ to $O(1)$. 

In the context of the arithmetic $P$ vs. $NP$ problem that we turn to next,
characterization by symmetries is  even more important. Because 
in this context we do not know how to use 
downward self-reducibility to prove any {\em any} flip theses. 
Specifically, the best result based on downward self reducibility 
for the usual nonuniform $P$ vs. $NP$ problem 
is the one in \cite{atserias}, which as we already discussed after 
Theorem~\ref{tflip}, does not efficiently yield a global obstruction set against
all circuits (i.e., cannot even satisfy F1 (a)).
This is akin to a  similar phenomenon that has already been observed in
complexity theory: namely, we know how to use random self reducibility to
reduce  worst case  hardness to average case  hardness in the context of the 
$\#P$ vs. $P$ problem, 
but not in the context of the $P$ vs. $NP$ problem, and indeed,
there is  compelling evidence \cite{trevisan,fort} that the usual reduction strategies 
based on self reducibility 
would not work in the  context of the $P$ vs. $NP$ problem.

\subsection{Proof of Theorem~\ref{tflip}}\label{sproofb}
For these reasons,  Theorem~\ref{tflip} 
proved in this  section is the main  result in the weak arithmetic setting.

The following is the  analogue of Proposition~\ref{prussell} in this case.

\begin{prop} \label{pcoRPE}
The problem of deciding if a given arithmetic circuit $C$ over $\Z$ 
computes $E(X)$  belongs to $co-RP$. 
\end{prop} 
\proof
For any $y \in \C$, and $i\not = j$, 
let $e_{ij}(y)$ denote an elementary $n\times n$ matrix with $1$'s on the diagonal,
$y$ in the $(i,j)$-th place, and zeroes everywhere else. 
By the proof of Guassian elimination, 
any matrix in $GL_n(\C)$ can be written as a product of 
elementary matrices, where by an elementary matrix we mean a matrix of the
form $e_{ij}(y)$, or a diagonal matrix, or  an elementary permutation matrix 
(that swaps some fixed two rows or columns). The total number of types of 
elementary matrices is clearly $O(n^2)$. 
Fix an explicit set $\{f_j\}$ of generators 
for the group $K$ (defined before the statement of Theorem~\ref{tE})
so that the total bit length of their description is $O(\poly(n))$.

By property (E) as per  Theorem~\ref{tE}, $C(X)=E(X)$ up to a nonzero constant 
multiple if and only if 
\begin{equation} \label{eqE0}
E(X)\not = 0
\end{equation}
identically as a polynomial, 
\begin{equation} \label{eqE1}
C(e X)=C(X), 
\end{equation}
for any elementary matrix $e$,
\begin{equation} \label{eqE2}
C(X)=C(X f_j), \quad \mbox{for all \ } j,
\end{equation}
and 
\begin{equation} \label{eqE3}
C(X)=0
\end{equation}
for any $X$  such that,
$X^1_i$, for each $i<m$,  is a vector with $1$ in the $i$-th location and
zero everywhere else, and the $m$-entry of $X^1_m$ is zero.
This last condition tests the property  E2 (2).

Testing if $C(X)$ satisfies (\ref{eqE0})--(\ref{eqE3}) is
an arithmetic circuit (polynomial) identity testing problem over $\Z$,
which belongs to co-RP. Specifically, to test (\ref{eqE0}) we choose $X$ randomly.
We need to test (\ref{eqE1}) separately
for each type of $e$. If $e$ is of the type $e_{ij}(y)$, we choose $y$ 
randomly and test (\ref{eqE1}) by choosing $X$ randomly. Similarly if $e$ is 
diagonal. If $e$ is an elementary permutation matrix, we just have to choose
$X$ randomly. Similarly for (\ref{eqE2}). For testing (\ref{eqE3}), we have 
choose $X$ randomly subject to the condition on $X$ specified there.
\qed

Testing (\ref{eqE1}) for a given elementary $e$ and  a given $X$ 
requires only $O(1)$ evaluations of the circuit $C$, and similarly for
(\ref{eqE0}), (\ref{eqE2}) and (\ref{eqE3}), just as in the case of
(\ref{eqpropp0}), (\ref{eqpropp1}) or (\ref{eqpropp2}).
The rest of the proof of Theorem~\ref{tflip}
is now like that of Lemma~\ref{lflip} using Proposition~\ref{pcoRPE} 
instead of Proposition~\ref{prussell}. 

For the proof of F2, the following conjecture plays the role of 
Conjecture~\ref{cstrongassumption1}.

\begin{conj} \label{cstrongassumption2}
Analogue of Conjecture~\ref{cstrongassumption1}
holds assuming that $S_{n,m}(s)$ is defined using 
(derandomization) of the algorithm in Proposition~\ref{pcoRPE} instead of the one in
Proposition~\ref{prussell}.
\end{conj} 

This finishes the proof of Theorem~\ref{tflip}.

\subsection{Derandomization of black box polynomial identity testing} \label{sderandom}
The proofs of Lemma~\ref{lflip}  and Theorem~\ref{tflip}  above also go through if
instead of assumming that $E$ does not have subexponential size 
circuits, we assume instead
that black box polynomial identity testing \cite{agrawal,russell}
can be derandomized. By this we mean that there exists 
a family  ${\cal H}=\cup_{n,m} H_{n,m}$ such that:

\begin{enumerate} 
\item {\em Short:} 
Each element $h$  of $H_{n,m}$ is a short hitting set \cite{agrawal} against 
all arithmetic circuits over $X=(x_1,\ldots,x_n)$  of bit size $\le m$. By a short 
hitting set $h$ we mean a set $\{X_1,\ldots,X_l\}$, $l=\poly(m)$, of inputs of total
bit size $O(\poly(n,m))$
such that for every circuit $C$ of total bit size
$\le m$ that computes a nonzero polynomial,
$h$ contains an input $X_C=X_i$, $i \le l$, such that $C(X_C) \not = 0$. 
\item {\em Rich:}   $H_{n,m}$ contains at least $2^{\Omega(m)}$ pairwise disjoint hitting sets.
\item {\em Easy to verify:} Given $n,m$ and $h$, whether $h \in H_{n,m}$ can be
verified in $\poly(n,m,\bitlength{h})$ time, where $\bitlength{h}$ denotes the
bit length of $h$.
\item {\em Easy to construct:} Given $n$ and $m$, a short $h \in H_{n,m}$ can 
be constructed in $\poly(n,m)$ time.
\end{enumerate}

The proof of Lemma~\ref{lflip} shows that this derandomization hypothesis holds if
$E$ does not have subexponential size circuits.
We leave the details of reworking the proofs of Lemma~\ref{lflip} and Theorem~\ref{tflip} 
with this less stringent derandomization hypothesis, instead of the assumption about $E$,
to the reader. No additional conjectures
such as Conjecture~\ref{cstrongassumption1} or \ref{cstrongassumption2} 
are needed in this case. 

Derandomization of black box polynomial identity testing is roughly 
equivalent to proving subexponential arithmetic circuit size 
lower bounds for multilinear functions in $E$; cf. Section 7.3  in \cite{russell}
and Section 5  in \cite{agrawal}. The notion of derandomization here is a bit 
stronger than that in \cite{agrawal,russell}. But the proofs there can be extended
to this stronger setting easily.

\subsection{Proofs of Lemma~\ref{lflipu} and Theorem~\ref{tflipu}} \label{suniform}
This  follows by uniformizing the proofs  of Lemma~\ref{lflip} and Theorem~\ref{tflip}. 
We omit the details.

\ignore{
\subsection{Proof of Lemma~\ref{lpvsnp}} \label{sprooflemmapvsnp}

Assuming $NP \not \subseteq P/poly$, \cite{atserias} gives a  probabilistic 
polynomial time algorithm for finding, given
any small circuit $C$, a counterexample on which it differs from SAT.
However, this algorithm is not nonadaptive (unlike the probabilistic algorithms
in the proofs of Propositions~\ref{prussell} and \ref{pcoRPE}), and hence, cannot be
used to prove Lemma~\ref{lpvsnp} along the lines of the proof of Theorem~\ref{tgeneric}.
Instead we use the probabilistic polynomial time algorithm  with
access to SAT oracle in \cite{fortnow} that produces a set $S_{n,m}$ 
of $\poly(m)$ satisfiable formulae over  $n$ variables $x_1,\ldots,x_n$
such that, for any circuit
$C$ with size $\le m$,  $S_{n,m}$ contains a counterexample $X^C=(x^C_1,\ldots,x^C_n)$
such that 
$C(X^C)\not = SAT(X^C)$. This algorithm  can be derandomized  assuming that
$E$ does not have subexponential size circuits with access to the SAT oracle.
Specifically, under this assumption, obvious extension of 
the (relativizable) proof technique in \cite{impa} 
gives a $\poly(m)$-time computable pseudorandom generator $g$ that takes a 
random seed of bitlength $O(\log m)$ and produces a pseudorandom sequence of length
$m$ that fools any circuit of size $m$ with access to the SAT oracle. 
We can now derandomize the algorithm above using this pseudorandom generator. 
The rest of the proof of Lemma~\ref{lpvsnp} is like that of Theorem~\ref{tgeneric}. 
The following conjecture is used in place of Conjecture~\ref{cstrongassumption1}
for proving F2.

\begin{conj} \label{cstrongassumption3}
Analogue of Conjecture~\ref{cstrongassumption1} 
holds for the pseudorandom generator as above.
\end{conj} 

We cannot prove F1 (b) though in the absence of any property like (E).

\qed

}

\section{Flip in the arithmetic setting}  \label{sarithproof}
In this section we prove Theorem~\ref{tflip2}. 

\subsection{Strong derandomization hypothesis} \label{sderanstrong}
We begin by specifying the strong derandomization hypothesis mentioned in the statement of
Theorem~\ref{tflip2}. It is a natural generalization of the derandomization hypothesis 
in the weak arithmetic setting described in Section~\ref{sderandom}. 

Let $C$ be an arithmetic circuit over $X=(x_1,\ldots,x_n)$ 
of size $\le m$. Let $S=[1,2^{m^2}]$ be the set of integers between $1$ and $2^{m^2}$ (say).
Since the degree of $C(X)$ is $\le 2^m$, by the standard  lemma \cite{schwarz},
the result of evaluating $C$ is nonzero with a high probability 
if $X$ is assigned a random element in $S^n$. It is critical here that the size of 
$S$ does not depend on the bitsize of the constants in $C$, since we are allowing arbitrary
constants from $\C$ in $C$. Indeed,  constants may not even have specifications of
finite bitlength if they are transcendental. Now we have a natural randomized polynomial 
time algorithm in the complex-RAM model for deciding if $C(X)$ is identically zero: 
(1) pick a random element in $S^n$, (2) evaluate $C(X)$, (3) say no if $C(X)$ is not zero, and
(4) yes otherwise. 
In the complex-RAM model  each memory location contains a complex number,
and each arithmetic operation ($+,-,*$) is unit-cost.
This is a black-box algorithm in the sense that it treats the circuit 
$C$ as a black-box subroutine. 

The derandomization hypothesis in the arithmetic setting is that this black box 
polynomial identity testing can be derandomized.  
By this we mean that there exists 
a family  ${\cal H}=\cup_{n,m} H_{n,m}$ such that:

\begin{enumerate} 
\item {\em Short:} 
Each element $h$  of $H_{n,m}$ is a short hitting set \cite{agrawal} against 
all arithmetic circuits over $X=(x_1,\ldots,x_n)$  of  size (rather than bit size) $\le m$.
By a short 
hitting set $h$ we mean a set $\{X_1,\ldots,X_l\}$, $l=\poly(n,m)$, of inputs of total
bit size $O(\poly(n,m))$ 
such that for every circuit $C$ of size
$\le m$ that computes a nonzero polynomial,
$h$ contains an input $X_C=X_i$, $i \le l$, such that $C(X_C) \not = 0$. 
\item {\em Rich:}   $H_{n,m}$ contains at least $2^{\Omega(m)}$ pairwise disjoint hitting sets.
\item {\em Easy to verify:} Given $n,m$ and $h$, whether $h \in H_{n,m}$ can be
verified in $\poly(n,m,\bitlength{h})$ time, where $\bitlength{h}$ denotes the
bit length of $h$.
\item {\em Easy to construct:} Given $n$ and $m$, a short $h \in H_{n,m}$ can 
be constructed in $\poly(n,m)$ time.
\end{enumerate}

\begin {lemma} 
The arithmetic derandomization hypothesis above holds assuming that $E$ does not
have subexponential size circuits, or less stringently, that analogous derandomization 
hypothesis holds over $F_p$, with the bitlength $\bitlength{p} = O(m^2)$, say.
\end{lemma} 

The derandomization hypothesis over $F_p$ is just like the arithmetic derandomization 
hypothesis above with the arithmetic circuits of size $\le m$ replaced by circuits over
$F_p$ of size $\le m$, and requiring each input in the hitting set to be over $F_p$ 
instead of $\Z$.

\proof
From the proof of Lemma~\ref{lflip} it follows (after appropriate modifications) that 
the derandomization hypothesis over $F_p$ holds assuming that $E$ does not have subexponential
size circuits.  Since each arithmetic circuit over $\Z$ corresponds to a circuit 
over $F_p$ obtained by reducing it modulo $F_p$,  the derandomization
hypothesis over $F_p$ implies the arithmetic derandomization hypothesis over $\Z$. \qed

The strong arithmetic derandomization hypothesis in
the strong arithmetic setting is obtained by letting $C(X)$ in the arithmetic  hypothesis above
be any function that can be approximated infinitesimally closely
by circuits of size $\le m$. Thus the hitting set is now against all functions that 
can be approximated infinitesimally closely by circuits of size $\le m$.

\subsection{Proof of Theorem~\ref{tflip2}} \label{sprooftflip2}
We now describe how to extend the proof of Theorem~\ref{tflip} to that 
of Theorem~\ref{tflip2}. We only consider 
Theorem~\ref{tflip2} (a),  since (b) is very similar.

The conditions F0-4 in Theorem~\ref{tflip2} can be proved just like those in
Theorem~\ref{tflip} in the weak arithmetic setting, letting the (strong) arithmetic 
hardness conjecture play the role of the weak arithmetic hardness conjecture, and 
letting the (strong) derandomization hypothesis above play the role of the weak 
derandomization hypothesis in Section~\ref{sderandom}. 

What remains to prove then is the property G. We turn to this next.

We follow the  terminology in the statement of the property G in 
Section~\ref{sarithflip}. 
Thus,  given $s \in {\cal O}_{n,m}$, $S_{n,m}(s)=\{X_1,\ldots,X_l\}$, $l=\poly(n,m)$,
denotes the small glbal obstruction set as in F1 (a). 
The space  $V$ is the space of 
polynomial functions in $X$ of degree $\le d=2^m$, and
$\Sigma_{n,m}$ is the set of the functions in $V$ that can be computed by 
arithmetic circuits of size $\le m$. 

Let $z$ be an additional homogenizing variable.
Given any $g(X) \in V$, let $g'(z,X)$ denote the homogeneous 
polynomial of degree precisely $d=2^m$ obtained from $g(X)$ by homogenizing it using $z$.
Let  $V'$ denote the space of homogenizations of the polynomials in $V$.
Let $\Sigma'_{n,m} \subseteq V'$ denote the set of all constant multiples of 
homogenizations of all  polynomials in $\Sigma_{n,m}$. This  set is homogeneous; i.e.
if $g'(z,X) \in \Sigma'_{n,m}$, then $a g'(z,X) \in \Sigma'_{n,m}$ for all $a \in \C$.
Let $\perm'(z,X)=z^{d-n} \perm(X) \in V'$ be the homogenization of $\perm(X)$.
Let $\psi'=\psi'_s: V' \rightarrow \C^l$ denote the homogeneous linear map such
that for any $g'(z,X) \in V'$, and any $i \le l$, 
\[ \psi'_s(g'(z,X))_i= g'(1,X_i).\] 
In other words, $\psi'_s(g'(z,X))$ is simply the $l$-tuple 
of evaluations of $g'(z,X)$ at various 
$X_i$'s, letting $z=1$, and $\psi'_s(g'(z,X))_i$ denotes the $i$-th entry in this tuple.

It is easy to show that any $g'(z,X) \in \Sigma'_{n,m}$ 
can be computed by an arithmetic circuit over $\C$ with
input $z$ and $X$ and of  size $\le m'= b m^2$ for some large enough constant $b$.
(The proof proceeds by induction on the depth of the circuit computing $g'(z,X)$.)
Hence it follows from
the strong arithmetic hardness conjecture for $\perm(X)$ that 
$\perm'(z,X)$ does not belong to the closure $\bar \Sigma'_{n,m}$ of 
$\bar \Sigma_{n,m}$ in the complex
topology. Assuming the strong derandomization hypothesis (cf. Section~\ref{sderanstrong}), 
it follows as in the proof of 
F1 in the strong arithmetic setting above, that, for any $s\in {\cal O}_{n,m'}$,
$S_{n,m'}(s)$ is also a global obstruction set against all functions in 
$\bar \Sigma'_{n,m}$. 
Specifically, this means that 
$\psi'_s(\perm'(z,X)) \not \in \psi'_s(\bar \Sigma'_{n,m})$. 
Replacing ${\cal O}_{n,m}$ by ${\cal O}_{n,m'}$ in the
obstruction family ${\cal O}$, we will assume, 
without of loss of generality, that, for any $s \in {\cal O}_{n,m}$, 
$S_{n,m}(s)$ is  a global obstruction set against all functions in $\bar \Sigma'_{n,m}$. 
This means
\begin{equation} \label{eqpsi}
\psi'_s(\perm'(z,X)) \not \in \psi'_s(\bar \Sigma'_{n,m})
\end{equation}
for any $s \in {\cal O}_{n,m}$.

Let $P(V')$ be the projective space of lines in $V'$ through the origin. Let
$P(\C^l)$ be the similar projective space associated with $\C^l$. Let $P(\bar \Sigma'_{n,m}) 
\subseteq P(V)$ denote the projective set associated with $\bar \Sigma'_{n,m}$. 
We can assume, without loss of generality that,  for any function $g'(z,X)$
in $\bar \Sigma'_{n,m}$, 
$S_{n,m}(s)$ contains a matrix $X_C$ such that $g'(1,X_C)\not = 0$; i.e., 
$\psi'_s(g'(z,X))$ is not an identically zero tuple. This is because the test 
for the property (P) also includes the test that the   function under consideration is 
not identically zero (cf. eq.(\ref{eqpropp0})),  and $S_{n,m}(s)$ is constructed on the
basis of the property (P).  
Thus  $\psi_s'$ gives a 
well defined map from $P(\bar \Sigma'_{n,m})$ to $P(\C^l)$. 
We denote this map by $\hat \psi_s'$.
We can also assume without loss of 
generality that each $S_{n,m}(s)$ contains an identity matrix. Since the permanent 
of the identity matrix is one, this means $\psi'_s(\perm'(z,X))$ is also not an 
identically zero tuple. 
We denote the point in $P(\C^l)$ corresponding to 
$\psi'_s(\perm'(z,X))$ by $\hat \psi'_s(\perm'(z,X))$.
Thus, by eq.(\ref{eqpsi}),
\begin{equation} \label{eqpsi2}
\hat \psi'_s(\perm'(z,X)) \not \in \hat \psi'_s(P(\bar \Sigma'_{n,m})) \subseteq P(\C^l)
\end{equation}
for any $s \in {\cal O}_{n,m}$.

To prove the property G for the permanent function,
it suffices to show that $\psi'_s(\perm'(z,X))$ does not belong to the closure of 
$\psi'_s(\bar \Sigma'_{n,m})$ in the complex topology.
This is equivalent to showing that $\hat \psi'_s(\perm'(z,X))$ does not belong to the 
closure of $\hat \psi'_s(P(\bar \Sigma'_{n,m}))$ in the complex topology.
By eq.(\ref{eqpsi2}), this follows from the following.

\begin{lemma} \label{lpiclosed}
The set $\hat \psi'_s(P(\bar \Sigma'_{n,m})) \subseteq P(\C^l)$ 
is already closed in the complex topology.
\end{lemma}

Fix $n,m$ and $s \in {\cal O}_{n,m}$. For simplicity, we drop the subscripts $s,n$ and $m$. 
Thus we denote $\hat \psi'_s$ by $\hat \psi'$, $\Sigma'_{n,m}$ by $\Sigma'$, and 
$\bar \Sigma'_{n,m}$ by $\bar \Sigma'$.

To prove lemma~\ref{lpiclosed}, we  need the following lemma.

\begin{lemma} \label{lpialg}
The set $\bar \Sigma' \subseteq V'$ 
is an algebraic variety (possibly reducible); i.e. the zero set of finitely many 
polynomials in the coordinates of $V'$. 
\end{lemma} 

\proof This follows from the following two facts from classical algebraic geometry: 

\noindent (1) The set $\Sigma \subseteq V$ and hence the set 
$\Sigma' \subseteq V'$ is a constructible set. (A set is called constructible 
(cf. Definition 2.30 in \cite{mumford}) if it can be expressed as a disjoint union 
$T_1 \cup \cdots \cup T_k$, where each $T_k = T_k'-T_k''$ for some algebraic variety
$T_k'$ and its subvariety $T_k''\subseteq T_k'$.)

This  can be proved  as follows. Fix an uninstantiated circuit $D$ of size $\le m$.
By an uninstantiated circuit, 
we mean the nodes of $D$ are labelled with the operators $+,-$ and $*$,
and the leaves are labelled by either the variables $x_i$'s, or constant parameters
$a_1,\ldots, a_j$, for some $j<m$. Clearly there are only finitely many uninstantiated 
circuits for given $m$. Fix any such $D$. 
Let $\Sigma_D\subseteq V$ be the set of all functions
that can be computed by some instantiation of $D$; i.e., by assigning
specific complex values to the constant parameters $a_1,\ldots, a_j$. Clearly, 
$\Sigma=\cup \Sigma_D$. So it suffices to show that $\Sigma_D$ is constructible. 
With $D$, we can associate an affine algebraic variety as follows. Associate a new variable 
$y_u$ with every 
internal node $u$  of $D$. (The leaves of $D$ are already associated with either variables
$x_i$'s or constant parameters $a_r$'s). Say the internal node $u$ is $*$, and $u_1$ and
$u_2$  are its children, possibly leaves. Then corresponding to $u$, we have an equation
$y_u=y_{u_1} * y_{u_2}$.  Let $\Pi_D$ denote the affine variety 
defined by all the equations associated with the internal nodes. 
Then $\Sigma_D$ is the projection  of $\Pi_D$ into $V$. (This corresponds to elimination
of all variables for the internal nodes and the parameters $a_1,\ldots,a_j$). Now (1) 
follows from  the fact (cf. Proposition 2.31  in \cite{mumford}) 
that the image of any affine variety under a regular (polynomial) map is a constructible 
set.   It need not be closed. See Chapter 2C in \cite{mumford} for the 
pathologies that can happen. This is the main problem that we have to deal with in the 
rest of the proof.

\noindent (2) The closure in the complex topology coincides with the closure in the
Zariski topology (cf. Theorem 2.33  in \cite{mumford}). 

Specifically this implies  the following. Since by (1), $\Sigma'$ is a constructible set, 
its closure $\bar \Sigma'$ in the complex topology is an algebraic variety (possibly 
reducible--we do not require a variety to be reducible in what follows).  \qed

Since  $\Sigma'$ is  homogeneous, its closure 
$\bar \Sigma' \subseteq V'$ is also homogeneous.  In conjunction with lemma~\ref{lpialg}, 
this means $\bar \Sigma'$ 
is a homogeneous algebraic subvariety of $V'$. Hence $P(\bar \Sigma')$ is a projective
subvariety of $P(V')$. 
Consider  the morphism  $\hat \psi'=\hat \psi'_s$ from $P(\bar \Sigma')$ to $P(\C^l)$ 
defined earlier.
To prove Lemma~\ref{lpiclosed}, it 
suffices to show that $\hat \psi'(P(\bar \Sigma'))$  is a  projective subvariety 
of $P(\C^l)$. This follows from 
the fact that the image of a morphism from a projective variety to another projective 
variety is closed (cf. Corollary 14.2 in \cite{eisenbud})--this is  a consequence
of the main theorem of elimination theory (cf. Theorem 14.1 in \cite{eisenbud}). 
This proves Lemma~\ref{lpiclosed}. 

Now the property  G follows. This proves  Theorem~\ref{tflip2} (a). 

\section{Implication in algebraic geometry} \label{simpli}
The algebraic variety $\bar \Sigma'_{n,m}$  associated above with the class of functions
computable by  small arithmetic circuits is rather wild and hard to study.  
The article GCT1  associates another variety with this class of functions. It is  called the 
{\em class variety} associated with the complexity class  $P$. Unlike $\bar \Sigma'_{n,m}$,
it has a natural action of the general linear group  $GL_{m^2}(\C)$. This  makes it possible
to study it using the techniques of geometric invariant theory \cite{mumfordfog}. 
The article GCT1 also associates similar class varieties with other complexity classes,
namely, $NC$, $NP$ and $\#P$.  Theorem~\ref{tflip2} implies 
that a formidable explicit construction problem
associated with these class varieties can be solved (in polynomial time)
assuming the strong arithmetic hardness
and derandomization hypotheses under consideration. To see this,
one simply has to rephrase Theorem~\ref{tflip2}  in terms of these varieties.
We  do it in this section for  the case of the 
strong arithmetic permanent  vs. determinant problem, the other cases being similar.

Towards that end, we first recall the class varieties associated by GCT1 with the 
complexity classes $NC$ and $\#P$. 
Let $Y$ be an $m\times m$ variable matrix. We think of its entries,
ordered say rowwise, as coordinates of ${\cal Y}=\C^r$, $r=m^2$.
Let $V=\C[Y]_m$ be the space of homogeneous polynomials  of degree $m$
in the variable entries of
$Y$. It is a representation of $G=GL({\cal Y})=GL_{r}(\C)$
with the following action. Given any $\sigma \in G$, map a polynomial  $g(Y) \in V$ 
to $g^\sigma(Y)=g(\sigma^{-1}(Y))$: 
\[ \sigma: g(Y) \longrightarrow g(\sigma^{-1} Y).\] 
Here $Y$ is thought of as an $m^2$-vector by straightening it rowwise.

Similarly, let $X$ be an $n\times n$ variable matrix, whose entries we think
of as coordinates of ${\cal X}=\C^{n^2}$ after ordering them rowwise.
Let $W=\C[X]_n$ be the space of forms (homogeneous polynomials) of degree $n$ in the
entries of  $X$.
It is  a representation of $H=GL({\cal X})=GL_{n^2}(\C)$.

Let $P(V)$ be the projective space of $V$ 
consisting  of the lines in $V$ through the origin. 
Let $P(W)$ be the projective space of $W$.
Identify $X$ with an $n\times n$ submatrix of $Y$, say, the bottom-right minor of $Y$, and
let $z$ be any variable entry of $Y$ outside $X$. We  use it as a homogenizing 
variable.
Define an embedding $\phi: W \hookrightarrow V$ by mapping 
any polynomial $h(X) \in W$ to $h^\phi(Y)=z^{m-n} h(X)$. This also defines an 
embedding of $P(W)$ in $P(V)$, which we denote by $\phi$ again.

Let $g=\det(Y)$, thought of  as a point in $P(V)$ (strictly speaking the line 
through $\det(Y)$ is a point in $P(V)$, but we ignore this distinction here).
Similarly, let $h=\perm(X) \in P(W)$, and 
$f=h^\phi=\perm^\phi(Y) \in P(V)$. 

Let 
\begin{equation}
\begin{array}{lclcl} 
\Delta_V[g,m]&=&\Delta_V[g]&=&\overline{G g} \subseteq P(V), \\
\Delta_W[h,n]&=&\Delta_W[h]&=&\overline{H h} \subseteq P(W), \\
\Delta_V[f,n,m]&=&\Delta_V[f]&=&\overline{G f} \subseteq P(V), \\
\end{array}
\end{equation}
where $ \overline {G g}$ denotes the projective closure of the
orbit $G g$ of $g$, and so on. Then, it follows from classical algebraic geometry
as in the proof of Lemma~\ref{lpialg} that  $\Delta_V[g,m]$ and $\Delta_V[f,m,n]$
are  projective varieties. Furthermore, it can be shown that
they are projective $G$-varieties, i.e.,
varieties with a natural action of $G$ induced by the action on the $G$-orbits. Similarly,
$\Delta_W[h,n]$ is a projective $H$-variety.
We call  $\Delta[f,n,m]$ the {\em class variety} 
of the complexity class $\#P$  since the permanent is $\#P$-complete \cite{valiant},
and $\Delta[g,m]$ the {\em class variety} of the complexity 
class $NC$ since the determinant belongs to $NC$ and is almost complete \cite{valiant}.

It is easy to show (cf. Propositions 4.1 and 4.4 in [GCT1]) that 
if $h=\perm(X)$ can be expressed linearly 
as the determinant of an $m\times m$ matrix, $m > n$, then 
\begin{equation} 
\Delta_V[f]=\Delta_V[f,n,m] \subseteq  \Delta_V[g,m]=\Delta_V[g],
\end{equation}
and conversely, if $\Delta_V[f,n,m] \subseteq \Delta_V[g, m]$,
then $f$ can be approximated 
infinitesimally closely by a point in $P(V)$ of the form
$\det(A Y)$, $A \in G$, thinking of $Y$ as an $m^2$-vector. 
The following conjecture is thus
equivalent to  the strong arithmetic permanent vs.
determinant conjecture  stated in Section~\ref{sarithmetic}. 

\begin{conj} (Strong arithmetic form of the permanent vs. determinant conjecture) [GCT1] 
\label{cpermvsdetstrong}
The point $f \in P(V)$ cannot be 
approximated infinitesimally closely as above  if $m=\poly(n)$, and more generally,
$m=2^{\log^a n}$ for any constant $a>0$. 

Equivalently, if $m=\poly(n)$, or more generally,
$m=2^{\log^a n}$, $a>0$ fixed, $n\rightarrow \infty$, then 
$\Delta_V[f,n,m] \not \subseteq \Delta_V[g,m]$. 
\end{conj} 

We now  restate  Theorem~\ref{tflip2} for this equivalent form of the
strong arithmetic permanent vs. determinant conjecture. 

An obstruction $s \in {\cal O}_{n,m}$
will now be against all points (functions) in $\Delta_V[g,m]$. Specifically, the
global obstruction set $S_{n,m}(s)=\{X_1,\ldots,X_l\}$, $l=\poly(n,m)$, will now have the
following property. Fix 
any homogeneous polynomial $p(Y)$ in $V$ that belongs to $\Delta_V[g,m]$ 
(thinking  of a homogeneous polynomial in $V$, by an abuse of notation, as a point in $P(V)$).
Then  there exists a counter example
$X_i \in S_{n,m}(s)$ such that 
$p'(X_i)\not = \perm(X_i)$, where $p'(X_i)$ is a polynomial obtained from $p(Y)$ by
substituting zero for all variables in $Y$ other than $z$ and $X$, substituting $1$ for $z$,
and $X_i$ for $X$.   
Equivalently, let $\psi=\psi_s: V \rightarrow \C^l$ be the homogeneous linear map 
that maps any homogeneous $p(Y) \in V$ to the point in
$\C^l$ corresponding to the tuple   $(p'(X_1,), \ldots,p'(X_l))$. 
As in the proof of Theorem~\ref{tflip2} in Section~\ref{sprooftflip2}, we can assume,
without loss of generality, that $\psi$ gives a well defined morphism from the 
projective variety $\Delta_V[g.m]$ to the projective variety $P(\C^l)$. We denote
this morphism by $\hat \psi=\hat \psi_s$. 
Its image is   $\hat \psi(\Delta_V[g,m]) \subseteq P(\C^l)$. 
We can also assume, as in  the proof of Theorem~\ref{tflip2} in Section~\ref{sprooftflip2},
that $\psi(f) \in \C^l$ is not an identically zero tuple. Hence it defines a point
in $P(\C^l)$, which we define by $\hat \psi(f)$.
Then that $S_{n,m}(s)$ is a global obstruction set is equivalent to saying that
$\hat \psi(f) \not \in \hat \psi(\Delta_V[g,m])$. The notion of an explicit proof and F0-4 
can now be formulated in this setting in the obvious manner; we omit the details.
Note that,
since  $\hat \psi$ is a well defined morphism from the projective variety
$\Delta_V[g,m]$ to the projective variety $P(\C^l)$, its image
$\hat \psi(\Delta_V[g,m])\subseteq P(\C^l)$ is already closed (projective subvariety) in
$P(\C^l)$ by the main theorem of elimination theory (cf. Corollary 14.2 in \cite{eisenbud}).
Hence the property G follows from F0-4 in this setting immediately by the main theorem of
elimination theory.

The following  is a restatement of  Theorem~\ref{tflip2} in this setting.

\begin{theorem}[Flip] \label{tflip3}
Assume Conjecture~\ref{cpermvsdetstrong}
and the strong arithmetic derandomization hypothesis (cf. Section~\ref{sderanstrong}).
Then Conjecture~\ref{cpermvsdetstrong}  has an explicit proof satisfying F0-4  and G as above.

More specifically, for any obstruction $s \in {\cal O}_{n,m}$, there is a linear map 
$\psi_s: V \rightarrow \C^l$ corresponding to the polynomial time computable 
global obstruction set 
$S_{n,m}(s)$ such that (1) it gives a well defined morphism $\hat \psi_s$ 
from  $\Delta_V[g,m]$ to $P(\C^l)$,  (2) $\psi_s(\Delta_V[g,m])$ is a closed projective
subvariety of $P(\C^l)$, and (3) 
$\hat \psi_s(f) \not \in \hat \psi_s(\Delta_V[g,m])$.  

Analogous result holds in the context of the strong arithmetic $P$ vs. $NP$ problem,
letting the similar  variety for the class $P$ defined in GCT1 
play the role of $\Delta_V[g,m]$ 
and letting the function $E(X)$ play the role of $\perm(X)$. 
\end{theorem}

We call the linear map $\hat \psi_s$ in Theorem~\ref{tflip3} 
an {\em explicit separator} between $\Delta_V[g,m]$ 
and $f=\perm^\phi(Y)$. We call it explicit  because,  given $s$, its specification
$S_{n,m}(s)$ can be computed in $O(\poly(n,m))$ time. We call $l=\poly(n,m)$
the {\em dimension}  of $\hat \psi_s$. 
Thus  Theorem~\ref{tflip3}  says that, assuming the strong arithmetic permanent vs. determinant
and derandomization conjectures,  one can construct 
an explicit family of linear separators of small dimension
between $\Delta_V[g,m]$ and $f=\perm^\phi(Y)$. 

It has to be stressed that Theorem~\ref{tflip3} critically depends on the exceptional 
nature of $f=\perm^\phi(Y)$ and $g=\det(Y)$. 
If one were to consider  general $f$ and $g$ in place of the permanent and  determinant,
the conclusion of Theorem~\ref{tflip3} will almost never hold.
For general $f$ and $g$, a global obstruction set $S_{n,m}$ that gives 
a  linear separator $\psi$ between $\Delta_V[g,m]$
and $f$ can be constructed (if it exists) by appropriately 
eliminating   $\dim(V)-r$ variables. This can be done  using  general purpose algorithms in 
algebraic geometry for computing multivariate resultants and Gr\"obner bases.
But these algorithms take $\Omega(\dim(V))$ space and $\Omega(2^{\dim(V)})$ time.
Since $\dim(V)$ is exponential in $n$ and $m$, 
the time taken  is at least double  exponential 
in $n$ and $m$, and the total bit length of $S_{n,m}$ is exponential in $n$ and $m$.
 Nothing better can be expected for  general $f$ and $g$, because elimination
theory is in general intractable. Specifically, the problem of computing the Gr\"obner 
basis is
EXPSPACE-complete \cite{mayr}. This  means  
it takes in general space that is exponential in the dimension of the
ambient space, which is  $P(V)$ here. 
In contrast, Theorem~\ref{tflip} says that  a short  specification $S_{n,m}$ 
of a  linear separator
between  $\Delta_V[g,m]$ and  $f=\perm^\phi(Y)$, can be computed in  $\poly(n,m)$ time
exploiting  the exceptional nature of $f$ and $g$. 
This may seem unbelievable.

At present,  such explicit separators of small dimension 
can be constructed  in algebraic geometry only between very 
special kinds of  algebraic varieties,  such as the Grassmanian or  the flag varieties 
\cite{fultonrepr}, and very special  kinds of points. This can be done using the 
second fundamental theorem of invariant theory \cite{fultonrepr,weyl} 
which gives a very nice  explicit set of generators for the ideals of these varieties.
But these  varieties have very low complexity
in comparison to $\Delta_V[n,m]$. For example, their complexity, according to a certain 
complexity measure on (quasi)-homogeneous spaces defined in \cite{vust}, is zero,
whereas that of $\Delta_V[g,m]$ is quadratic in $m$. Furthermore, they are normal,
whereas $\Delta_V[g,m]$ is not normal according to a recent result \cite{shrawan}.
The problem of explicit construction of linear  separators when the 
underlying variety is not normal and its complexity  is  so high seems  very formidable
and far beyond the reach of the existing machinery in algebraic geometry.
Theorem~\ref{tflip3} says that
such formidable explicit construction problems in algebraic geometry are hidden 
underneath the hardness and derandomization hypotheses in complexity theory. 


\section{Flip in the boolean setting} \label{sboolean}
To get an efficient pseudorandom generator,  it does not suffice to just assume that 
$P \not = NP$. One needs a stronger average case assumption, namely, existence of one way functions.
Similarly, to get a flip theorem in the context of the usual (boolean) $NP \not \subseteq P/poly$ 
conjecture, one needs to assume a  stronger average case 
form of this boolean conjecture based on characterization by symmetries. In this section 
we state this conjecture (Conjecture~\ref{cboolean}). 
The corresponding flip theorem (Theorem~\ref{tboolean}) then follows as a direct corollary of the 
main result in \cite{impa} on derandomization of BPP.

We begin with a preliminary motivating result in the context of the following strengthening 
of Conjecture~\ref{cpvsnpzero}. 

\begin{conj}\label{cpvsnpvariant}
Analogues of Conjecture~\ref{cpvsnpzero} and  Conjecture~\ref{cstrongpvsnp} 
hold for any integral  nonzero 
$e(X)$ with the  properties E1' and E2 as  in Theorem~\ref{tE} (b). 
\end{conj} 

This gives a purely group-theoretic definition of hardness in the context of the arithmetic
$P$ vs. $NP$ problem.

\begin{theorem}[Flip for property E1] \label{tgenericE1}
Analogues of Theorems~\ref{tflip} and \ref{tflip2} 
hold for any nonzero integral $e(X)$ with the  properties
E1' and E2 (as  in Theorem~\ref{tE} (b)). 
\end{theorem}

This is proved just like Theorems~\ref{tflip} and \ref{tflip2},
with  Conjecture~\ref{cpvsnpvariant}  playing the role of 
Conjecture~\ref{cpvsnpzero} and the property E1' the role  of   E1.

Now we turn to the boolean setting. The  following is a stronger form  of 
the $NP \not \subseteq P/poly$ Conjecture.

Let $S$ be the set integers of bit length at most $n^3$ (say). 
Let $C$ be a boolean circuit whose input is the  bit specification 
of $X$ with  entries  in $S$. 
Let $A$ be a co-RP algorithm for testing if $C$  has properties E1' and E2 
akin to the algorithm in the proof of Proposition~\ref{pcoRPE} (for testing E1 and E2)
with the following difference. 
Whenever we used a random number  in that algorithm, we use  a random integer of bitlength at most
$n^3/3$, and
instead of standard generators of $GL_n(\C)$, we now use standard generators of $SL_n(\C)$.

\begin{conj}[Stronger invariant theoretic average case form of the  $NP \not \subseteq P/poly$ conjecture]
\label{cboolean}
Let $C$ be any  boolean circuit   of $\poly(n)$ bit size whose input is bit specification  of $X$ 
with entries in $S$.
Suppose $C$  comes with 
a {\em promise} that 
$\mbox{prob}\{C(X)=0\}$, $X\in S$, is small, say $<1/n$, where $C(X)$ denotes
the boolean function computed by $C$. 

Then the algorithm $A$ above for testing if $C$ has properties E1' and E2 says NO 
with high probability ($\ge 1/poly(n)$).
\end{conj}

The promise is necessary in Conjecture~\ref{cboolean} since there exist small circuits 
with the properties E1' and E2   that are zero almost everywhere but not everywhere.

\begin{prop} 
Conjecture~\ref{cboolean}  implies $NP \not \subseteq P/poly$.
\end{prop}
\proof 
Let $E_b(X)$ be the boolean function which is zero if $E(X)$ is zero and one otherwise.
Clearly $E_b(X)$ has  properties  E1' and E2. Furthermore, 
computation of $E_b(X)$  is $NP$-complete \cite{gurvits}. Hence 
it suffices  to show that any boolean circuit computing $E_b(X)$
satisfies  the promise. But the number of zeros of $E_b(X)$ 
is the same as those of $E(X)$. Hence by the Schwarz-Zippel lemma, 
$\mbox{prob}\{E_b(X)=0\}$, $X\in S$, is bounded by $\deg(E(X))/|S|= n^{kn^2}/2^{n^3} < 1/n$. \qed

Conjecture~\ref{cboolean} 
basically says that the symmetries E1' and E2 of $E_b(X)$ are hard to approximate  on the average. 
This is an invariant theoretic average case form of the worst case assumption that 
$E_b(X)$ is hard to compute (as expected since it is  $NP$-complete).
It will be interesting to study the relationship (if any) between this average case 
assumption and the standard average case assumptions in complexity theory, such as existence
of one way functions.

\begin{theorem}[Flip in the boolean setting] \label{tboolean}
Suppose Conjecture~\ref{cboolean}
holds and also that the complexity class $E$ does not have subexponential size
circuits (or less stringently, that the co-RP algorithm $A$ above 
can be derandomized in a black box fashion
very much as in Section~\ref{sderandom}).

Then for every $n$ and $m=\poly(n)$,
it is possible to compute in $\poly(n,m)=\poly(n)$ time 
a small set $S_{n,m}=\{X_1,\ldots,X_r\}$,  $r=\poly(n,m)=\poly(n)$, 
of $n\times n$  matrices with entries in $S$ such that for every  boolean  circuit 
$C$  satisfying the promise in Conjecture~\ref{cboolean} and with total bit size 
$\le m$ (and hence, in particular, for any boolean circuit of size $\le m$ claiming to compute $E_b(X)$),
$S_{n,m}$ contains a matrix $X_C$ which is a counter example against
$C$ (as detected in the algorithm $A$).

Furthermore, assuming an appropriate  stronger form
(analogous to Conjecture~\ref{cstrongassumption2})
of the assumption that $E$ does not have subexponential size circuits,
(or less stringently, that the co-RP algorithm $A$ above 
can be derandomized in a black box fashion)
Conjecture~\ref{cboolean}, and hence, $NP \not \subseteq P/poly$ conjecture,
has an  explicit proof--i.e., there exists an obstruction family 
$\tilde {\cal O}$ satisfying F0-F4--except that the obstructions are now only against
small circuits satisfying the promise in Conjecture~\ref{cboolean}.
\end{theorem}

The  new ingradient here is formulation of Conjecture~\ref{cboolean}, i.e.,
formulation of the conjecturally correct nonadaptive co-RP algorithm algorithm $A$ 
for finding a counterexample against any small boolean circuit claiming to compute $E_b(X)$.
Once this is done, Theorem~\ref{tboolean}  is just a  direct corrollary of the 
main result in \cite{impa} on derandomization of BPP,
because the algorithm A can be derandomized under standard assumptions therein.
This algorithm  $A$  is to be  contrasted with  the {\em adaptive} probabilistic polynomial
time algorithm in \cite{atserias} for finding a counterexample against a small boolean circuit
claiming to compute SAT, assuming $NP \not\subseteq P/poly$.

Let us finish this section with one more variant of a flip theorem.

\begin{theorem}[Flip over a finite field] \label{tflipfinite}
Analogue of Lemma~\ref{lflip}  holds
over a large enough finite field $F_p$, $p > 2n$ (say), 
 instead of $\Q$ or  $\C$, provided  
in the definitions of F1-4 we confine ourselves to the circuits  
with the promise that the polynomials computed by them  have the same 
degree as that of  $\perm(X)$ (otherwise the circuit cannot compute $\perm(X)$ 
for trivial reasons). 

Similar analogue of Theorem~\ref{tflip} holds for hardness of the  function $E(X)$ 
over a large enough $F_p$ as in Conjecture~\ref{cstrongpvsnp}. 
\end{theorem}

This is also   proved  like Theorem~\ref{tflip}. 

\section{Rigidity} \label{sproofc}
The proof (cf. Section~\ref{sproofb})  of the  flip Theorem~\ref{tflip} 
works for any  function $e(X)$, $X=(x_1,\ldots,x_n)$, over $\Q$ 
in the  place  of  $E(X)$ as long as $e(X)$ has the following properties:

\noindent {\bf (1)} It is characterized by symmetries in the following sense:

\begin{defn} \label{dcharsym}
 We say that $e(X)$ is  characterized by symmetries 
if it is the only nonzero  polynomial (up to a constant multiple) with rational
coefficients  that satisfies a small
($\poly(n)$) number of algebraic polynomial identities with integral 
coefficients  (in the spirit of those in the property (E)),
each having a specification of $\poly(n)$ bitlength and containing $O(1)$ terms.
Here each identity is of the form 
\[ g(e(Y_1),\ldots, e(Y_k))=0,\] 
where $g(u_1,\ldots,u_k)$ is a  polynomial computable by a circuit over $\Z$ of
$O(1)$ size with input $u_i$'s, and each
$Y_i$ can be computed by a circuit over $\Z$ of $\poly(n)$ bit size with input $X$. 

If  we  only  require that 
each  $g(u_1,\ldots,u_k)$  be computable by  a $\poly(n)$ bit size circuit over $\Z$ with
input $u_i$'s, we say that $e(X)$ is weakly
characterized by symmetries.
\end{defn} 

The circuits specifying the identities   here can be nonuniform. 

\noindent {\bf (2)} $e(X)$ cannot be computed by an arithmetic circuit over $\Q$ 
of $\poly(n)$ bit size. 

Here (1) implies that
there is a nonadaptive co-RP/poly algorithm  for deciding if a given arithmetic
circuit $C$ computes $e(X)$ (akin to that in the proof of Proposition~\ref{pcoRPE}), 
where a co-RP/poly algorithm 
 means a nonuniform algorithm in the form of a $\poly(n)$ size 
circuit with random advice in addition to the usual input. Nonuniformity has 
to be allowed since the circuits specifying the identities in Definition~\ref{dcharsym} 
can be nonuniform. It is easy to
see that the proof of the  flip theorem goes through even in the presence of
such nonuniformity. It also goes through even 
when  $e(X)$ is required to be 
characterized by symmetries  in a weaker sense,  except that F1 (b)
need not hold in this weaker setting.

\begin{prop} \label{prigidity}
The number of $e(X)$ over $\Q$ 
that are  characterized by symmetries in a weaker sense 
(Definition~\ref{dcharsym})
is $\le 2^{\poly(n)}$.
\end{prop}
\proof This  holds 
because the total bit length of the specification of the identities in
Definition~\ref{dcharsym} in terms of small circuits is $O(\poly(n))$. \qed

The proposition 
implies that  the  proof technique of the  flip Theorem~\ref{tflip},
which only works for  functions with  properties (1) and (2),  is {\em extremely rigid}. 
By this  we mean that it only works for $2^{\poly(n)}$ number of functions in place of
$e(X)$. This is also the case for  Flip Theorem~\ref{tboolean} in the 
boolean setting.

This form of rigidity is extremely severe in comparison 
to the {\em mild rigidity} constraint that the natural proof barrier \cite{rudich} places 
\footnote{Ignoring the constructivity condition in \cite{rudich}}
on proof techniques for the $NP \not \subseteq P/poly$ conjecture: namely, that 
they  should work for less than $2^N/poly(N)$ number of functions, where $N=2^{n}$ is 
the size of the truth-table specification of an $n$-ary boolean function.

It is a {\em plausible}
that any proof   of the arithmetic or boolean $P$ vs. $NP$ conjecture (or any of the 
related conjectures under consideration in this paper) has to be 
extremely rigid. This is because by
Theorem~\ref{tflip} any proof of the (weak) arithmetic $P$ vs. $NP$ 
conjecture is close to an explicit proof.
But the explicitness condition seems  so severe that any proof that comes even close to 
an explicit proof may work for only rare exceptional functions (like the permanent or $E(X)$).
That is, just mildly rigidity which suffices to bypass the 
the natural proof barrier \cite{rudich} may not be enough, and a proof 
may be forced to be extremely rigid, like that Theorem~\ref{tflip} or \ref{tboolean}.

\ignore{In contrast, a proof technique has to be only mildly rigid to bypass the
natural proof barrier in \cite{rudich}

Here  extreme rigidity of a proof technique for the arithmetic or the boolean 
nonuniform $P$ vs. $NP$ conjecture is defined as follows.
First let us consider the boolean case. 
Let us assume that the proof technique has 
provided  as in 
the natural proof paper \cite{rudich} 
a formal statement 
of a Useful Property ($UP_n$), which lies at its heart, and assuming which 
it plans to prove $NP \not \subseteq P/poly$. Formally, $UP_n$ is a subset of the set
of all $n$-ary boolean functions. Its definition is 
as in \cite{rudich} and we do not restate it here. 
Let $N=2^n$ be the truth table size of specifying any $n$-ary boolean 
function.

\begin{defn} \label{drigidcomplexity}
We say that a proof technique for the $NP \not \subseteq P/poly$ conjecture 
is 
\begin{enumerate} 
\item 
{\em Nonrigid (or probabilistic)} if  $|UP_n| \ge 2^N/N^c$, for some constant $c>0$,
\item {\em Mildly rigid} if $|UP_n| \le 2^N /N^c$ for every constant $c>0$,
as $n\rightarrow \infty$, or more strongly, 
if  $|UP_n| \le 2^N/2^{n^a}$, for some constant $a\ge 1$,
\item {\em Rigid} if  $ |UP_n| \le 2^{\epsilon N}$,
for some $0 \le \epsilon < 1$, 
\item {\em Strongly rigid} if 
$|UP_n| \le 2^{N^\epsilon}$, for some $0 \le \epsilon < 1$,  and 
\item {\em Extremely rigid} if 
$|UP_n| \le 2^{\poly(n)}$, i.e., $\le 2^{n^a}$ for some constant $a>0$.
\end{enumerate}
\end{defn} 

Thus probabilistic proof techniques to which the natural proof
barrier \cite{rudich} applies are nonrigid, and 
if a proof technique is mildly rigid it bypasses this
barrier, since it violates the largeness criterion in \cite{rudich}. 
The  definition    extremely rigidity in the arithmetic setting is similar.

}

\subsubsection*{Acknowledgement:} 
The author is grateful to Janos Simon and Josh Grochow for helpful discussions.
\end{document}